\documentclass[reprint,amsmath,amssymb, aps, prx,superscriptaddress]{revtex4-1}

\usepackage{graphicx}
\usepackage{epsfig}
\usepackage{color}
\usepackage{amsmath}
\usepackage{amsfonts}
\usepackage{mathrsfs}
\usepackage{txfonts}
\usepackage{color}
\usepackage{subfig}

\usepackage[breaklinks]{hyperref} 
\hypersetup{colorlinks=true, linkcolor=blue, citecolor=blue, filecolor=blue, urlcolor=blue}

\newcommand{\bst}{\mathcal{T}}
\newcommand{\bea}{\begin{eqnarray}}
\newcommand{\eea}{\end{eqnarray}}

\begin{document}
	\title{Detecting symmetry fractionalization in gapped quantum spin liquids by magnetic impurities}
	\author{Shuangyuan Lu}
	\affiliation{Department of Physics, The Ohio State University, Columbus OH 43210, USA}
	\author{Yuan-Ming Lu}
	\affiliation{Department of Physics, The Ohio State University, Columbus OH 43210, USA}
	\date{\today}
	
	\begin{abstract}
    We study the Kondo effect of spin-$1/2$ magnetic impurities in gapped $Z_2$ spin liquids on two-dimensional lattices. We find that if the impurity is placed at a high-symmetry location, a nontrivial spinon fractionalization class of the impurity site symmetry group will necessarily lead to a non-Kramers doublet in the Kondo screening regime, protected by associated crystalline symmetries. This is in sharp contrast to a featureless screening phase in the usual Kondo effect. We demonstrate this symmetry-protected topological degeneracy by an exactly solvable model and by the large-$N$ theory. Based on this effect, we discuss how thermodynamic measurements in the limit of dilute magnetic impurities can be used to detect symmetry fractionalization in gapped $Z_2$ spin liquids. 
    \end{abstract}
	\maketitle

	\section{Introduction}
	Quantum spin liquids (QSLs)~\cite{Balents2010,Savary2016,Zhou2017} have attracted much interests in the past few decades due to its exotic properties transcending the Landau scheme of symmetry breaking. In particular, the presence of anyons which obey fractional statistics\cite{Wilczek1990Book}, is among the most exciting manifestations of the topological order and long-range entanglement in QSLs\cite{Wen2002,Wen2007Book}, with potential applications in topological quantum computations~\cite{Nayak2008}. A number of QSL candidate materials with various crystalline symmetries have been discovered experimentally~\cite{Kanoda2022,Zhou2017,Wen2019,Takagi2019}.
	
	Meanwhile, there is a gap between theoretical diagnosis and experimental measurements to identify QSLs. On one hand, various theoretically computable quantities has been proposed to sharply characterize topological orders, such as the topological entanglement entropy\cite{Kitaev2006EE,Levin2006} and modular matrices\cite{KESKI-VAKKURI2022,Zhang2012}. On the other hand, most existing experiments aim at ruling out long-range orders in the low temperature, deterred by the difficulty of directly probing unique features of QSLs. In particular, compared to gapless $U(1)$ spin liquids with clear signatures in inelastic neutron scattering (INS)~\cite{Benton2012} or thermal transport~\cite{Kanoda2022,Zhou2017}, a gapped symmetric QSL is more featureless and harder to detect experimentally. While the long-range entanglement and fractional statistics, as a definitive character of topological orders, is difficult to access experimentally, the fractional symmetry quantum number\cite{Laughlin1999} (formally known as symmetry fractionalization\cite{Wen2002,Essin2013,Tarantino2016,Barkeshli_2019,Chen2017}) of anyons provide extra features to characterize and identify the topological order, which is usually easier to probe experimentally. In the well known example of fractional quantum Hall effects (FQHEs), indeed the fractional charge is experimentally observed in the nineties, much earlier than the recently confirmed fractional statistics\cite{Feldman2021}. One question arises naturally: can symmetry fractionalization be experimentally detected as a direct evidence for a QSL state? 
	

	For gapped QSLs, which is the focus of this work, there are two major challenges to experimental detection of fractionalization. First of all, the fractionalized excitations such as spinons are charge neutral, therefore insensitive to charge transport probes which played a crucial role in identifying fractionalization in FQHEs\cite{Feldman2021}. Secondly, a gapped symmetric QSL usually has no features both in the bulk and on the boundary, making it very hard to access experimentally. This is unlike the $U(1)$ spin liquids, whose gapless excitations can be probed by INS in the case of emergent photons\cite{Benton2012}, or thermal transport in the case of spinon Fermi surfaces\cite{Kanoda2022,Zhou2017}. Is it possible to experimentally identify a gapped QSL? Previously, INS spectroscopy has been proposed to exhibit features of fractional statistics\cite{Morampudi2017} and spinon symmetry fractionalization\cite{Wen2002,Essin2014}. In this work, we look into magnetic impurities and Kondo effects in gapped QSLs, and show that they can provide distinct thermodynamic signatures of symmetry frationalization in QSLs, in the Kondo screening regime.   
	
	The Kondo effect in QSLs has previously been studied both in theories \cite{Khaliullin1995,Dhochak_2010,Doretto_2009,Ribeiro2011,Vojta_2016,Das_2016,Kolezhuk2006,Hu2022} and in experiments \cite{Yamamoto_2018,Gomilsek2019,Chen2022}, focusing on gapless QSLs.
 	In this paper, we explore the Kondo effect in gapped $Z_2$ QSLs. Similar to the distinctions between Kondo effects in metals and in insulators (with a vanishing density of states), the Kondo effect in gapped $Z_2$ QSLs differs qualitatively from gapless QSLs. In particular, due to the energy gap for spinon excitations, there is a finite threshold of Kondo coupling strength to screen the magnetic impurity\cite{Withoff1990,Satori1992,Saso1992,Itoh1993,Takegahara1993,Chen1998}. Most remarkably, we find that when a half-integer-spin impurity is placed at a high-symmetry location in the crystal hosting a gapped $Z_2$ QSL, the Kondo screening phase will feature a non-Kramers doublet localized at the impurity site, protected by fractionalized crystalline symmetries in the $Z_2$ QSL. This symmetry protected degeneracy lead to distinct signatures in the thermodynamics, such as specific heat, which can serve as ``smoking gun'' evidence for symmetry fractionalization in a gapped QSL. This phenomena is demonstrated by an exactly solvable model and large-$N$ parton mean-field theory, which agree with each other.


	\section{Main results}
	
\begin{figure}
    \centering
    \includegraphics[width=1\columnwidth]{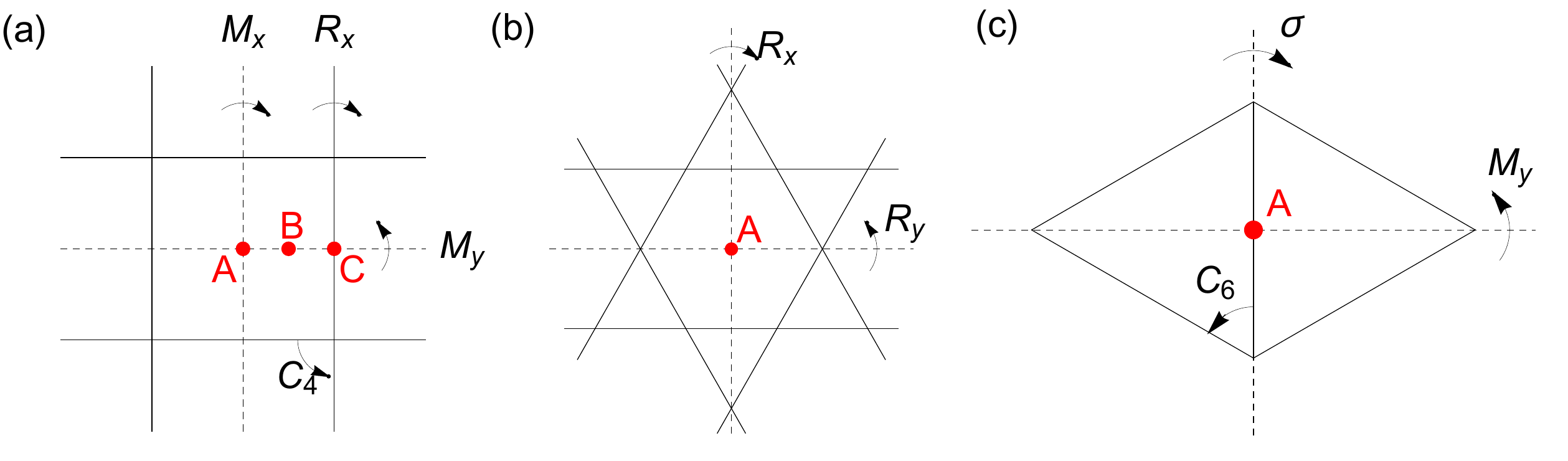}
    \caption{High-symmetry impurity sites that can be used to detect the symmetry fractionalization of spinons in a gapped symmetric $Z_2$ spin liquid, on the (a) square, (b) kagome and (c) triangular lattices in two dimensions.}
    \label{fig:2}
\end{figure}
	
We first present the major results in Fig. \ref{fig:2} and TABLEs \ref{table.1}-\ref{table.3}. Consider a gapped symmetric $Z_2$ spin liquid on a two-dimensional lattice, whose Hilbert space consists of a spin-$1/2$ (or a Kramers doublet) on each lattice site. A spin-$1/2$ magnetic impurity located at certain high-symmetry position of the lattice can be used to diagnose the symmetry fractionalization class\cite{Essin2013,Barkeshli_2019,Tarantino2016} of the $Z_2$ spin liquid phase. Specifically, when such a magnetic impurity is coupled symmetrically to the $Z_2$ spin liquid, in the Kondo screening regime, there may or may not be a two-fold degeneracy (a non-Kramers doublet) protected by the crystalline symmetry of the impurity site, depending on the fractionalization class of spinons in the $Z_2$ spin liquid. 

Fig. \ref{fig:2} illustrates the three lattices enumerated in this work, i.e. the square, kagome and triangular lattices. In the presence of $SO(3)$ spin rotational symmetry, the classification of symmetric $Z_2$ spin liquids on these lattices\cite{Lu_2018,Qi2018} are summarized in TABLE \ref{table.1}-\ref{table.3}. A part of the symmetry fractionalization data can be detected by presence/absence of non-Kramers doublets for Kondo-screened magnetic impurities located at different high-symmetry sites, such as a plaquette center, a nearest-neighbor link center, or on a mirror plane. 

\begin{table}[h]
    \centering
    \begin{tabular}{|c|c|c|c|c|}
    \hline
      Algebraic identity&$\omega\in\mathcal{H}^2(G,\mathcal{A})$&$\omega^e$\cite{Yang2016}&$\omega^\epsilon$\cite{Wen2002}&Impurity site\\
      \hline
      $(R_x)^2$&$\omega_{R_x,R_x}$&$(-1)^{p_4}$&$\eta_\sigma$&-\\
    \hline
    $(M_y)^2$&$\omega_{M_y,M_y}$&$(-1)^{p_3+p_4}$&$\eta_\sigma\eta_{xpx}$&-
    \\
    \hline
$(C_4R_x)^2$&$\omega_{C_4R_x,C_4R_x}$&$(-1)^{p_4+p_7}$&$\eta_\sigma\eta_{\sigma C_4}$&-\\
    \hline
    $M_xM_yM_x^{-1}M_y^{-1}$&$\frac{\omega_{M_x,M_y}}{\omega_{M_y,M_x}}$&$(-1)^{p_1}$&$\eta_{xy}$&A \\
        \hline
    $(M_y\mathcal{T})^2$&$\omega_{M_y\mathcal{T},M_y\mathcal{T}}$&$(-1)^{p_3+p_8+1}$&$-\eta_t\eta_{xpx}$&B\\
        \hline
        $R_xM_yR_x^{-1}M_y^{-1}$&$\frac{\omega_{R_x,M_y}}{\omega_{M_y,R_x}}$&$(-1)^{p_2}$&$\eta_{xpy}$&C\\
    \hline
    \end{tabular}
    \caption{All gapped $Z_2$ spin liquids of spin-$1/2$'s on the square lattice ($G=p4m\times Z_2^\bst$), characterized by $2^6$ fractionalization classes\cite{Lu_2018,Qi2018}, and their realizations in the Schwinger boson\cite{Yang2016} and Abrikosov fermion\cite{Wen2002} representations. 3 of the 6 independent $Z_2$ invariants can be detected by magnetic impurities located at A, B and C sites in Fig. \ref{fig:2}(a). }
    \label{table.1}
\end{table}	

\begin{table}[h]
    \centering
    \begin{tabular}{|c|c|c|c|c|}
    \hline
      Algebraic identity&$\omega\in\mathcal{H}^2(G,\mathcal{A})$&$\omega^e$\cite{Yang2016}&$\omega^\epsilon$\cite{Wen2002}&Impurity site\\
      \hline
      $(R_x)^2$&$\omega_{R_x,R_x}$&$(-1)^{p_2+p_3}$&$\eta_\sigma$&-\\
    \hline
    $(R_y)^2$&$\omega_{R_y,R_y}$&$(-1)^{p_2}$&$\eta_\sigma\eta_{\sigma C_6}$&-
    \\
    \hline
    $R_xR_yR_x^{-1}R_y^{-1}$&$\frac{\omega_{R_x,R_y}}{\omega_{R_y,R_x}}$&$(-1)^{p_1}$&$\eta_{12}$&A \\
        \hline
    \end{tabular}
    \caption{All gapped $Z_2$ spin liquids of spin-$1/2$'s on the kagome lattice ($G=p6mm\times Z_2^\bst$), characterized by $2^3$ fractionalization classes\cite{Lu_2018,Qi2018}, and their realizations in the Schwinger boson\cite{Wang2006} and Abrikosov fermion\cite{Lu2011} representations. 1 of the 3 independent $Z_2$ invariants can be detected by magnetic impurities located at A site in Fig. \ref{fig:2}(b). }
    \label{table.2}
\end{table}	

\begin{table}[h]
    \centering
    \begin{tabular}{|c|c|c|c|c|}
    \hline
      Algebraic identity&$\omega\in\mathcal{H}^2(G,\mathcal{A})$&$\omega^e$\cite{Yang2016}&$\omega^\epsilon$\cite{Wen2002}&Impurity site\\
      \hline
      $\sigma^2$&$\omega_{\sigma,\sigma}$&$(-1)^{p_2}$&$\eta_\sigma$&-\\
    \hline
    $(M_y)^2$&$\omega_{M_y,M_y}$&$(-1)^{p_2+p_3}$&$\eta_{\sigma C_6}$&-
    \\
    \hline
    $\sigma M_y\sigma^{-1}M_y^{-1}$&$\frac{\omega_{\sigma,M_y}}{\omega_{M_y,\sigma}}$&$(-1)^{p_1}$&$\eta_{12}$&A\\
        \hline
    \end{tabular}
    \caption{All gapped $Z_2$ spin liquids of spin-$1/2$'s on the triangular lattice ($G=p6mm\times Z_2^\bst$), characterized by $2^3$ fractionalization classes\cite{Lu_2018,Qi2018}, and their realizations in the Schwinger boson\cite{Wang2006} and Abrikosov fermion\cite{Lu2016} representations. 1 of the 3 independent $Z_2$ invariants can be detected by magnetic impurities located at A site in Fig. \ref{fig:2}(c). }
    \label{table.3}
\end{table}	

In a gapped system with a vanishing density of states, a finite Kondo coupling strength is required to enter the Kondo screening regime\cite{Withoff1990,Satori1992,Saso1992,Itoh1993,Takegahara1993,Chen1998}, where the system typically reaches a featureless paramagnetic ground state. The proposed non-Kramers doublet protected by crystalline symmetries at the impurity site of the $Z_2$ spin liquid is therefore a striking and unusual phenomenon. Below we describe the physical picture behind this observation. 

A gapped $Z_2$ spin liquid hosts three types of anyons (or superselection sectors): bosonic spinon $e$, vison $m$, and their bound state $\epsilon=e\times m$ known as a fermionic spinon\cite{Kitaev_2003}. In a symmetric $Z_2$ spin liquid on a lattice with an odd number of spin-$1/2$'s in each unit cell, spinons $e$ and $\epsilon$ must carry spin-$1/2$ each, while vison $m$ is spinless\cite{Zaletel2015}. As a result, when a spin-$1/2$ magnetic impurity is coupled to such a $Z_2$ spin liquid,
to reach a spin-singlet ground state, it can only be screened by a spinon $e$ or $\epsilon$. The same conclusion holds if we replace spin-$1/2$ by a Kramers doublet with $\bst^2=-1$ in the argument.

In the presence of crystalline and time reversal symmetries, different gapped $Z_2$ spin liquids are distinguished by their symmetry fractionalization classes, classified by 2nd group cohomology $\mathcal{H}^2(G,\mathcal{A})$, where the symmetry group is $G=SG\times Z_2^\bst$ ($SG$ being the space group), and $\mathcal{A}=Z_2\times Z_2$ is the fusion group of Abelian anyons in the $Z_2$ spin liquid\cite{Essin2013}. Thanks to the $SO(3)$ spin rotational symmetry, the vison fractionalization class is uniquely fixed on the three lattices\cite{Qi2015,Lu_2017}, leading to the classification shown in Table \ref{table.1}-\ref{table.3}\cite{Qi2018,Lu_2018}, characterized by projective representations of $G$ carried by spinons. Consequently, the singlet bound state formed by the impurity spin and the screening cloud of spinon can carry a projective representation of the impurity site symmetry group. This leads to the non-Kramers degeneracy at the impurity site, protected by both crystalline and time reversal symmetries. 

There are two types of impurity sites of particular interests to this work: (i) the impurity site lies at the intersection of two mirror planes, such as site A and C in Fig.\ref{fig:2}; (ii) the impurity site lies on a mirror plane, such as site B in Fig. \ref{fig:2}. In case (i), since the two mirror symmetries $M_x$ and $M_y$ commute in the Hilbert space of the impurity spin, if they anticommute on the spinon screening the impurity spin (i.e. $\omega_{M_x,M_y}/\omega_{M_y,M_x}=-1$), the bound state of impurity spin-$1/2$ and spinon will carry a projective representation of the site symmetry group $G_s=Z_2^{M_x}\times Z_2^{M_y}$, leading to a 2-fold degeneracy protected by the mirror symmetries. On the other hand, if the two mirror actions commute on the spinon (i.e. $\omega_{M_x,M_y}/\omega_{M_y,M_x}=+1$), the bound state will carry a linear representation of the site symmetry and hence no degeneracy in the screening regime. In case (ii), the impurity spin-$1/2$ on a mirror ($M_x$) plane carries a projective representation $(M_x\bst)^2=-1$. If the spinon screening the impurity carry a linear representation with $\omega_{M_x\bst,M_x\bst}=+1$, their bound state forms a projective representation of the site symmetry group $G_s=Z_2^{M_x}\times Z_2^\bst$, leading to a 2-fold non-Kramers degeneracy protected by both mirror $M_x$ and time reversal symmetries. In contrary, if the spinon carries a projective representation with $\omega_{M_x\bst,M_x\bst}=-1$, the bound state would instead form a linear representation $(M_x\bst)^2=+1$ with no degeneracy in the screened phase. This shows how the response of a $Z_2$ spin liquid to impurity spin-$1/2$'s can diagnose a part of the spinon fractionalization data in the spin liquid phase, as summarized in Table \ref{table.1}-\ref{table.3}. 

\begin{figure}[h]
    \centering
    \includegraphics[width=0.48\columnwidth]{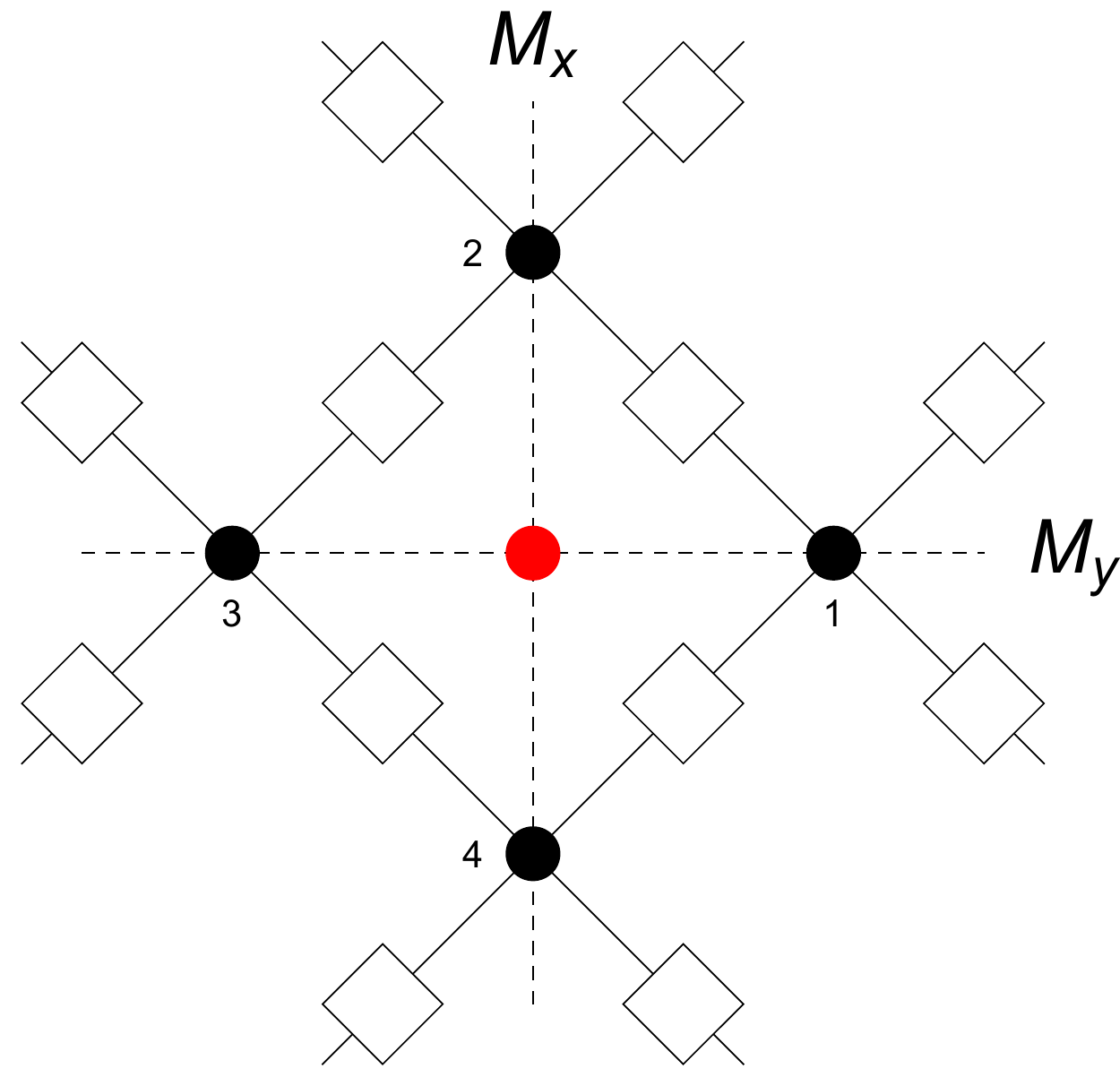}
    \caption{An exactly solvable model for a spin-$1/2$ impurity in a $Z_2$ spin liquid with $SU(2)$ symmetry. The squares on link centers are qubits in the toric code\cite{Kitaev_2003}. Each black dot on a star/vertex represents a Hilbert space of spin $0\oplus 1/2$. The red dot denotes the spin-$1/2$ impurity with a site symmetry $G_s=Z_2^{M_x}\times Z_2^{M_y}$\cite{supp}.}
    \label{fig:solvable model}
\end{figure}

\section{Methods}

We use two methods to demonstrate the symmetry-protected non-Kramers degeneracy induced by a spin-$1/2$ impurity in $Z_2$ spin liquids, in the Kondo screening regime. We focus on case (i) of impurity site symmetry $G_s=Z_2^{M_x}\times Z_2^{M_y}$, where the impurity spin-$1/2$ lies at the intersection of two mirror planes $M_{x,y}$. First we construct an exactly solvable model by modifying the toric code, to show the exact degeneracy protected by two mirror symmetries. Next we use the large-$N$ approach to solve the Kondo problem in a symmetric $Z_2$ spin liquid, and to compute the temperature dependence of thermodynamic quantities. 

First we present an exactly solvable model illustrated in Fig. \ref{fig:solvable model}. The bulk $Z_2$ spin liquid is described by:
\begin{equation}
\begin{aligned}
    &\hat{H}_{bulk}=-\sum_s A_s -\sum_p B_p\\ 
    &-\sum_s \Delta\big[\frac{(A_s+1)}2 P_s(S=0)+\frac{(1-A_s)}2 P_s(S=1/2)\big]
\end{aligned}\label{eq:model}
\end{equation}
where $A_s$ and $B_p$ are the star and plaquette operators in Kitaev's toric code\cite{Kitaev_2003}. In addition to one qubits on each link, there is a 3-dimensional Hilbert space of spin $0\oplus\frac12$ on each site/vertex (see Fig. \ref{fig:solvable model}). In the limit $\Delta\gg1$, each $e$ and $\epsilon$ particle will carry spin-$1/2$ of the site/vertex Hilbert space, while $m$ particles are spinless. The fractionalization class associated with the impurity site symmetry group $G_s=Z_2^{M_x}\times Z_2^{M_y}$ is given by\cite{Essin2013} 
\bea\label{eq:model frac class}
\frac{\omega^e_{M_x,M_y}}{\omega^e_{M_y,M_x}}=\frac{\omega^m_{M_x,M_y}}{\omega^m_{M_y,M_x}}=+1,~~~\frac{\omega^\epsilon_{M_x,M_y}}{\omega^\epsilon_{M_y,M_x}}=-1.
\eea
Next we introduce a spin-$1/2$ impurity located at the center of plaquette $(1234)$ in Fig. \ref{fig:solvable model}, which is coupled to the bulk spin liquid as follows:
\begin{equation}
    H_{imp} = J \sum_{i=1}^4 \vec{S}_i \cdot \vec{S}_{imp}+E_c(A_1+A_2+A_3+A_4-3)^2+\Delta_\epsilon B_p[1234]
\end{equation}
In addition to the usual Kondo coupling $J$, we also introduce (i) a Coulomb repulsion $E_c$ for spinons, which makes sure the impurity is screened by one spinon, and (ii) an energy $\Delta_\epsilon$ coupled to the plaquette opeartor on plaquette $[1234]$, to control which type of spinons ($e$ vs. $\epsilon$) will screen the impurity spin. Assuming $E_c\gg1,J$, the Kondo screening regime happens when $J>4/3$, leading to 4 degenerate states in the low energy manifold where a single spinon is located at one neighboring site (out of $1,2,3,4$). When $\Delta_\epsilon<1$, the bosonic spinons cost less energy and will screen the impurity, and the 4-fold degenerate can be completely lifted with a unique paramagnetic ground state without breaking any symmetry\cite{supp}. This is consistent with the trivial fractionalization class of $e$ particles in (\ref{eq:model frac class}). When $\Delta_\epsilon>1$, however, the fermionic spinons cost lower energy and are responsible for the Kondo screening. As detailed in supplemental materials\cite{supp}, the 4-dimensional low energy space can be split into two doublets, each of which form an irreducible projective representation of the impurity site symmetry group $G_s=Z_2^{M_x}\times Z_2^{M_y}$. As a result, a 2-fold degeneracy protected by two mirror symmetries $M_{x,y}$ will emerge in the Kondo screening regime, as indicated by the nontrivial fractionalization class (\ref{eq:model frac class}) of $\epsilon$ particles. Therefore we have demonstrated the correspondence between nontrivial fractionalization class of spinons screening the impurity, and protected 2-fold degeneracy in the Kondo screening regime. 

Next we use a large-$N$ mean field theory to solve the Kondo problem in symmetric $Z_2$ spin liquids. Both the bulk and impurity spins are represented by fermonic partons with $Sp(2N)$ symmetry\cite{ran2006continuous}:
\bea
S^{ab+}=\frac12(c^{a\dagger}_\uparrow c_{\downarrow}^{b}+c^{b\dagger}_\uparrow c_{\downarrow}^{a}),~~~S^{ab,z}=\frac12(c^{a\dagger}_\uparrow c_{\uparrow}^{b}-c^{b\dagger}_\downarrow c_{\downarrow}^{a})
\eea
with $1\leq a,b\leq N$. They reduce to the familiar $SU(2)$ spin-$1/2$ case when $N=1$. The model consists of a $Z_2$ spin liquid in the bulk described by parton mean-field ansatz of $Sp(2N)$ partons
\begin{equation}\label{Eq.bulk}
    H_{bulk}= \sum_{a=1}^N\sum_{i,j} J_{i,j}\psi_{i}^{a,\dagger} u_{i,j} \psi^a_j+h.c. 
\end{equation}
where we denote $\psi_i^a=(c_{i,\uparrow}^a,c^{a\dagger}_{i,\downarrow})^T$, and the Kondo coupling between a $Sp(2N)$ impurity spin and its neighboring spins:
\begin{equation}\label{Eq.imp}
H_{imp}=\sum_{\langle j,imp \rangle} \frac{J}{N} \mathbf{S}_{j}^{a b} \cdot \mathbf{S}_{imp}^{b a}+\frac{J^\prime}{N^3} (\mathbf{S}_{j}^{a b} \cdot \mathbf{S}_{imp}^{b a})^2
\end{equation}
As detailed in supplemental materials\cite{supp}, the bulk parton ansatz can be exactly realized in solvable models in analogy to Kitaev's honeycomb model\cite{Kitaev2006}, and choosing different link parameters $\{u_{ij}\}$ can lead to either trivial or nontrivial fractionalization classes for fermionic spinons $\{\psi_i^a\}$, with $M_xM_yM_x^{-1}M_y^{-1}=\pm1$. A self-consistent mean-field calculation, which becomes exact in the large $N$ limit, reveals a Kondo screening phase separated from the unscreened phase by a Kondo temperature $T_K(J)$, for Kondo couplings beyond a finite threshold $J>J_c$\cite{supp}. In the Kondo screening regime with $T<T_K(J)$, while the trivial fractionalization class ($M_xM_yM_x^{-1}M_y^{-1}=+1$) shows a unique paramagnetic ground state, the nontrivial fractionalization class ($M_xM_yM_x^{-1}M_y^{-1}=-1$) exhibits two degenerate ground states which cannot be mixed by any local perturbations preserving mirror symmetries $M_{x,y}$\cite{supp}. This again demonstrated our conclusion, that a nontrivial spinon fractionalization class will lead to symmetry protected zero modes localized at high-symmetry impurity sites.

 \begin{figure}
    \centering
    \includegraphics[width=\columnwidth]{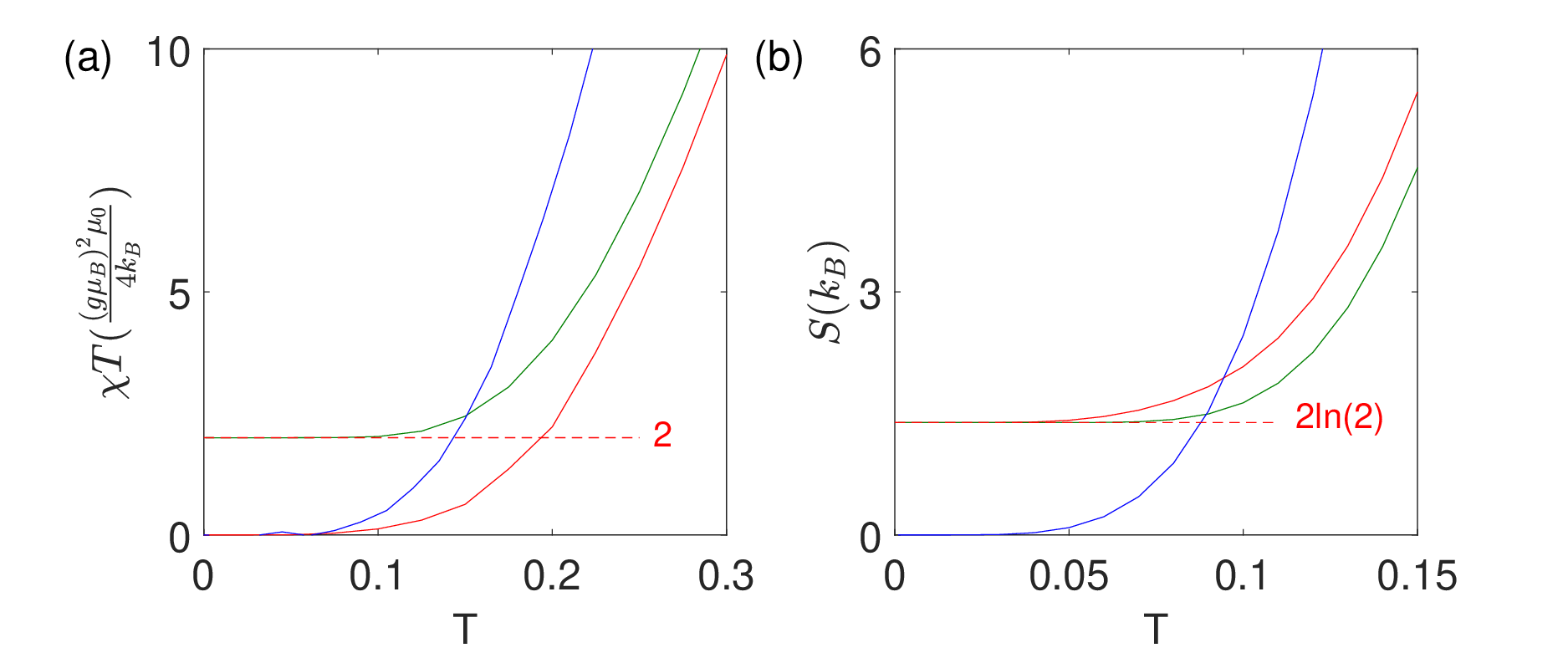}
    \caption{ The temperature dependence of (a) uniform magnetic susceptibility $\chi(T)$ and (b) entropy $S(T)$ contributed by Kondo impurities in different regimes: the unscreened regime of free moments at the impurity sites (green), the Kondo screening regime in $Z_2$ spin liquids with a trivial (blue) vs. nontrivial (red) spinon fractionalization class. The calculations are performed for two distant impurities with site symmetry $G_s=Z_2^{M_x}\times Z_2^{M_y}$ on a $20\times20$ lattice\cite{supp}.}
    \label{fig:thermodynamics}
\end{figure}

\section{Experimental implications}

The large-$N$ mean field theory also allows us to predict distinct experimental signatures of the anomalous Kondo screening phase described above. The temperature dependence of uniform magnetic susceptibility $\chi(T)$ and the entropy $S(T)=\int_0^{T}\frac{C_v(t)}{t}\text{d}t$ are shown in Fig. \ref{fig:thermodynamics}. The shown impurity contribution shown in the figure susceptibility $\chi(T)$ and specific heat $C_v(T)$, by subtracting the bulk contribution of Hamiltonian $H_{bulk}$ from the total amount of $H_{bulk}+H_{imp}$. Three different regimes can be differentiated from each other by inspecting the susceptibility and entropy (by integrating the specific heat) at low temperatures: (i) In the unscreened regime, the magnetic impurity behaves as a free moment, leading to $\chi(T)\sim1/T$ and a finite entropy of $k_B\ln2$ per impurity, colored green in Fig. \ref{fig:thermodynamics}. (ii) In the Kondo screening regime, for $Z_2$ spin liquid with trivial fractionalization class, Kondo screening leaves a unique paramagnetic ground state below the Kondo temperature, and therefore exponentially decaying thermodynamic responses $\chi(T),C_v(T)\sim e^{-\Delta/k_BT}$, as colored blue in Fig. \ref{fig:thermodynamics}. (iii) A $Z_2$ spin liquid with a nontrivial fractionalization class, on the other hand, features a symmetry protected non-Kramers doublet (2-fold degeneracy) localized at each impurity site in the Kondo screening regime. As a result, while the susceptiblity vanishes exponentially at low temperatures $\chi(T)\sim e^{-\Delta/k_BT}$, there is a low energy entropy of $k_B\ln2$ per impurity below the Kondo temperature, as colored red in Fig. \ref{fig:thermodynamics}.
The sharp differences between the three scenarios provide clear experimental features to detect a gapped $Z_2$ spin liquid with a nontrivial symmetry fractionalization class for spinons. 

Theoretically we only discussed the case of isolated impurities in the large-$N$ self-consistent mean-field theory described earlier. In real materials, we expect our predictions in Fig. \ref{fig:thermodynamics} to hold in the case of \emph{dilute magnetic impurities}, where the average distance $r$ between neighboring impurities is much larger than the bulk correlation length $\xi$ of the gapped spin liquid. In this case, the splitting of the symmetry protected zero modes localized at impurity sites will be small $\sim Je^{-C_0r/\xi}$, where $C_0$ is a constant of order one. This will lead to a peak in specific heat at low temperature $T\sim Je^{-C_0r/\xi}$.

\section{Concluding remarks}

In this work, we show that when magnetic impurities with half integer spins are coupled to symmetric $Z_2$ spin liquids in an isotropic magnet with $SU(2)$ symmetry, a spinon will form a singlet bound state with the impurity in the Kondo screening regime. This bound state will feature a symmetry protected 2-fold degeneracy, if the spinon fractionalization class of the impurity site symmetry is nontrivial, therefore leading to a non-Kramers doublet localized at the impurity site. We further show that this local degeneracy in the Kondo screening regime can be distinguished from other scenarios by the low temperature behaviors of magnetic susceptibiltiy and specific heat, hence unveiling a new way to detect symmetry fractionalization in QSLs. 

In the future, it will be insightful to apply the new angle proposed in this work to examine the candidate materials of QSLs, where magnetic impurities are known to exist at high symmetry sites e.g. in Herbertsmithite\cite{Fu2015}. Theoretically, this work also provides a new idea to probe fractionalization in models of strongly correlated electrons, by studying the impurity problem e.g. using numerical methods. 

\acknowledgements{We thank Biao Huang for discussions and related collaborations at an early stage of this work. This work is supported by National Science Foundation (NSF) under award number DMR 1653769 (YML), by Center for Emergent Materials at The Ohio State University, a NSF MRSEC through NSF Award No. DMR-2011876 (SL).}

\bibliography{kondo_SL}

\clearpage

\onecolumngrid

\begin{center}
 {\bf Supplemental materials}   
\end{center}

\appendix

\section{A brief review of symmetry fractionalization in $Z_2$ spin liquids}

Below we provide a brief summary on the classification of gapped symmetric $Z_2$ spin liquids on the square lattice, with a single spin-$1/2$ on each site. 

Most generally, given a topological order described by a unitary modular tensor category $\mathcal{C}$, a $G$-symmetry enriched topological (SET) phase is mathematically classified by symmetry action $\rho:~G\rightarrow Aut(\mathcal{C})$, and twisted group cohomology $\mathcal{H}^2_\rho(G,\mathcal{A})$ where $\mathcal{A}$ represents all Abelian anyons in category $\mathcal{C}$\cite{Essin2013,Barkeshli_2019,Tarantino2016}. In the special case where symmetries do not permute different anyons in $\mathcal{C}$, different SET phases are classified by group cohomology $\mathcal{H}^2(G,\mathcal{A})$, which characterizes the symmetry fractionalization in the SET phase\cite{Essin2013}. Physically, the localized symmetry $\{U_g|g\in G\}$ on the anyons form a projective representation\cite{Wen2002} of the symmetry group $G$:
\bea
U_gU_h=\omega_{g,h}U_{g,h},~~~\omega_{g,h}\in\mathcal{A}.
\eea
Acting on anyon $a$, the product of $U_g$ and $U_h$ actions differ from $U_{gh}$ action by a $U(1)$ phase given by the braiding phase between anyon $a$ and Abelion anyon $\omega_{g,h}$:
\bea
\omega_{g,h}^a=\frac{S_{a,\omega_{g,h}}}{S_{a,0}}\in U(1)
\eea
where $S_{a,b}$ is the modular S matrix of topological order $\mathcal{C}$.

\subsection{Square lattice}

\begin{figure}
    \centering
    \includegraphics[width=0.4\columnwidth]{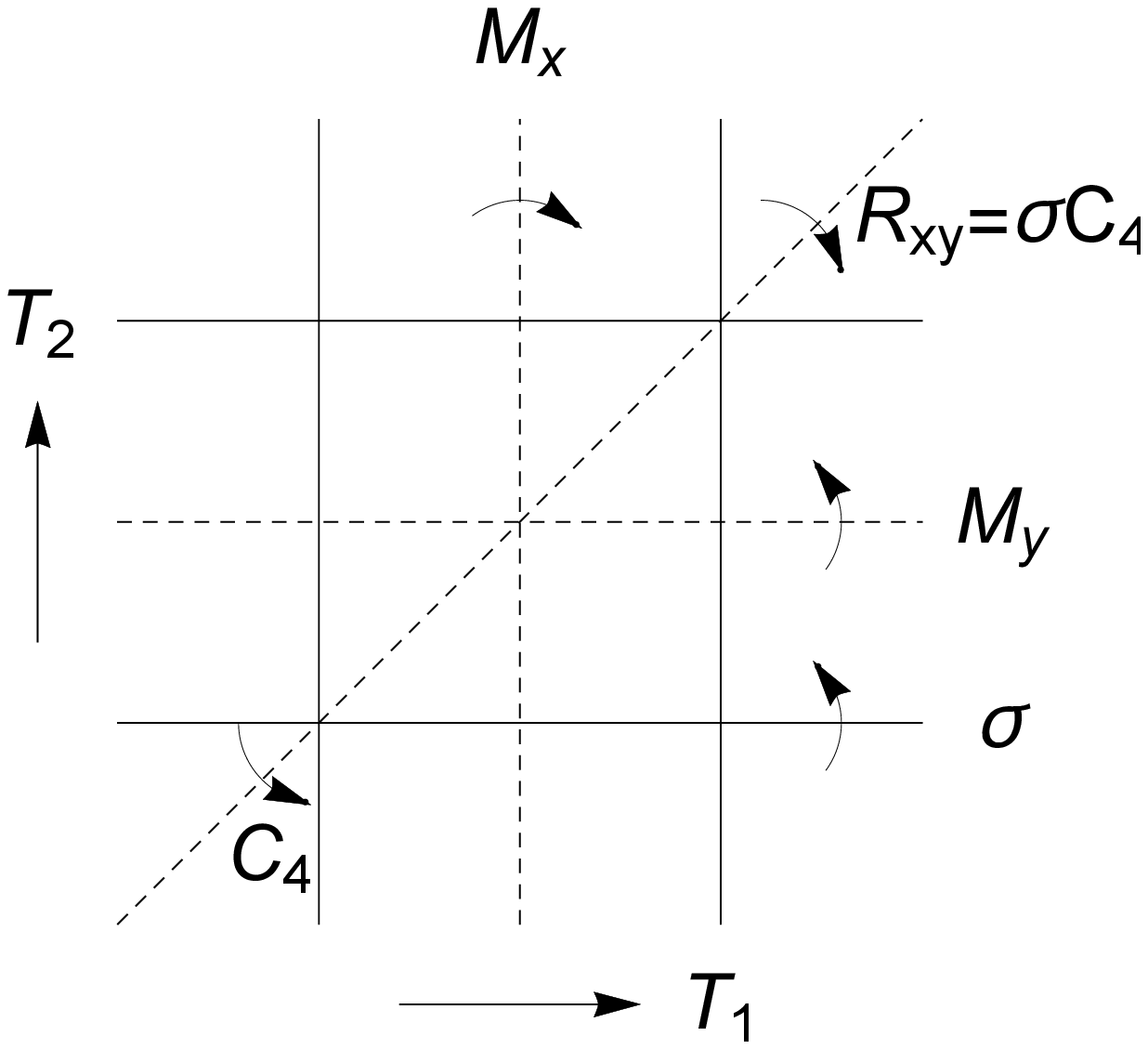}
    \caption{Space group symmetries of the square lattice.}
    \label{fig:square}
\end{figure}

As illustrated in Fig. \ref{fig:square}, the space group symmetry $p4m$ of the square lattice is generated by translation $T_1$, site-centered 4-fold rotation $C_4$ and mirror reflection $\sigma$. All other symmetry elements can be generated by them, for example in Fig. \ref{fig:square}
\begin{eqnarray}
 &T_2=C_4T_1C_4^{-1},\\
&M_y=T_2\sigma,\\
&R_{xy}=\sigma C_4,\\
&M_x=C_4M_yC_4^{-1}.
\end{eqnarray}
The full symmetry group is the product of space group symmetry $p4m$, time reversal symmetry $Z_2^{\mathcal{T}}$, and the spin rotational symmetry $SO(3)$. On a square lattice of spin-$1/2$'s, the symmetry action of spin rotations is fixed by the symmetry and Hilbert space in a relatively simple fashion: $e$ and $\epsilon$ particles each carry spin-$1/2$, while $m$ particles carry spin-$0$. We consider the rest of the symmetry group $G=p4m\times Z_2^\bst$, where the symmetry fractionalization classes of $Z_2$ spin liquids are classified by 2nd group cohomology
\bea
\mathcal{H}^2(G,\mathcal{A}=Z_2\times Z_2)=\big[\mathcal{H}^2(G,Z_2)\big]^2
\eea
Here $\mathcal{A}$ is the fusion of Abelian anyons in a $Z_2$ spin liquid. Each one of the two $[\omega]\in\mathcal{H}^2(G,Z_2)$ classes can be viewed as the classification for symmetry fractionalization on a specific anyon, for example $[\omega^e]$ for bosonic spinon $e$ and $[\omega^\epsilon]$ for fermionic spinon $\epsilon$. The associated symmetry fractionalization class for the vison $m=e\times\epsilon$ can be obtained by the fusion rule:
\bea\label{eq:fusion rule}
\omega^m_{g,h}=\omega^e_{g,h}\omega^\epsilon_{g,h}\omega_t(g,h),~~~\forall~g,h\in G.
\eea
where $[\omega_t(g,h)=\pm1]$ are a set of twist factors determined by the symmetry group and the underlying topological order.  

As shown in Table \ref{tab:classification}, the different cohomology classes of $\mathcal{H}^2(G,Z_2)=(Z_2)^{10}$ are characterized by 10 different $Z_2$-valued invariants\cite{Essin2013}. Naively, there will be 20 independent $Z_2$ invariants, 9 for bosonic spinons $[\omega^e]$ and 10 for fermionic spinons $[\omega^\epsilon]$. Luckily, as we will see below, the $SO(3)$ spin rotational symmetry and spin-$1/2$ Hilbert space fix 14 of them and reduce everything to only 6 independent $Z_2$ invariants\cite{Lu_2017}, summarized in Table \ref{tab:classification}.

\begin{table}[h]
    \centering
    \begin{tabular}{|c|c|c|c|c|c|}
    \hline
      Algebraic identities&$[\omega]\in\mathcal{H}^2(G,\mathcal{A})$&$\omega^e$\cite{Yang2016}&$\omega^\epsilon$\cite{Wen2002}&$\omega^m$\cite{Qi2015}&Twist factor $\omega_t$\cite{Essin2013,Lu_2017}\\
      \hline 
      $\bst^2$&$\omega_{\bst,\bst}$&$-1$&$-1$&1&1\\
      \hline 
      $T_1T_2T_1^{-1}T_2^{-1}$&$\frac{\omega_{T_1,T_2}}{\omega_{T_2,T_1}}$&$(-1)^{p_1}$&$\eta_{xy}$&-1&1\\
      \hline
      $\sigma^2$&$\omega_{\sigma,\sigma}$&$(-1)^{p_4}$&$\eta_\sigma$&1&-1\\
    \hline
      $(\sigma\bst)^2$&$\omega_{\sigma\bst,\sigma\bst}$&$-1$&$-\eta_\sigma\eta_{\sigma\bst}\equiv-1$&1&1\\
    \hline
    $(M_y)^2$&$\omega_{M_y,M_y}$&$(-1)^{p_3+p_4}$&$\eta_\sigma\eta_{xpx}$&1&-1\\
    \hline
    $(M_y\mathcal{T})^2$&$\omega_{M_y\mathcal{T},M_y\mathcal{T}}$&$(-1)^{p_3+p_8+1}$&$-\eta_t\eta_{xpx}\eta_\sigma\eta_{\sigma\bst}\equiv-\eta_t\eta_{xpx}$&1&1\\
        \hline    
$(C_4\sigma)^2$&$\omega_{C_4\sigma,C_4\sigma}$&$(-1)^{p_4+p_7}$&$\eta_\sigma\eta_{\sigma C_4}$&1&-1\\
    \hline
    $(C_4\sigma\bst)^2$&$\omega_{C_4\sigma\bst,C_4\sigma\bst}$&$-1$&$-\eta_\sigma\eta_{\sigma\bst}\eta_{\sigma C_4}\eta_{C_4\bst}\equiv-1$&1&1\\
    \hline
        $M_x\sigma M_x^{-1}\sigma^{-1}$&$\frac{\omega_{M_x,\sigma}}{\omega_{\sigma,M_x}}$&$(-1)^{p_2}$&$\eta_{xpy}$&1&-1\\
    \hline
    $(C_4^2\bst)^2$&$\omega_{(C_4)^2\bst,(C_4)^2\bst}$&$-1$&$-\eta_{C_4}\equiv-1$&-1&-1\\
    \hline
    \end{tabular}
    \caption{The classification of gapped $Z_2$ spin liquids of spin-$1/2$'s on the square lattice, characterized by 6 independent $Z_2$-valued invariants\cite{Yang2016,Lu_2018}, and their realizations in the Schwinger boson\cite{Yang2016} and Abrikosov fermion\cite{Wen2002} representations.}
    \label{tab:classification}
\end{table}

First, similar to the $SO(3)$ spin rotational symmetry, the time reversal symmetry action is fixed by the Hilbert space: $e$ and $\epsilon$ are Kramers doublets with $\bst^2=-1$, while each $m$ is a spin-0 particle with $\bst^2=+1$\cite{Essin2013}. Secondly, the $SO(3)$ spin rotational symmetry completely fix the vison fractionalization class $[\omega^m]$\cite{Qi2015}, as shown in Table \ref{tab:classification}. This means only the fermionic spinon fractionalization class $[\omega^\epsilon]$ is labeled by independent invariants, and it fully determines the bosonic spinon fractionalization class $[\omega^e]$ by fusion rule (\ref{eq:fusion rule}). Finally, for a gapped $Z_2$ spin liquid, the spin-$1/2$ Hilbert space per site further fixes $(M\bst)^2=\omega_{M\bst,M\bst}\equiv-1$ on spinons $e$ and $\epsilon$, as long as the reflection plane $M$ crosses an odd number of sites per unit cell\cite{Lu_2018,Qi2018}. A similar conclusion can be made for a 2-fold rotation (or inversion) $I_s$ centered on a lattice site (with a spin-$1/2$ on it): $(I_s\bst)^2=\omega_{I_s\bst,I_s\bst}\equiv-1$ for an arbitrary anyon carrying spin-$1/2$\cite{Lu_2018,Qi2018}. For fermionic spinons, this means $\eta_\sigma\equiv\eta_{\sigma\bst}$, $\eta_{\sigma C_4}\equiv\eta_{C_4\bst}$, and $\eta_{C_4}\equiv1$. Therefore we reach the 6 independent $Z_2$ invariants\cite{Lu_2018} summarized in Table \ref{table.1}.   

Below we make a few comments on the relation between different algebraic identities in the cohomology class, to elaborate on how Table \ref{table.1} is obtained from Table \ref{tab:classification}. First, for elements $\omega\in\mathcal{H}^2(G,Z_2)$, we have
\bea
\omega_{(C_4)^2\bst,(C_4)^2\bst}\equiv\omega_{(C_4)^2,(C_4)^2}\omega_{\bst,\bst}
\eea
hence the identity $(C_4)^4=e$ does not give rise to an extra invariant beyond Table \ref{tab:classification}. For spinons, this also indicates that
\bea
\omega^{e,\epsilon}_{(C_4)^2,(C_4)^2}\equiv1
\eea
Another algebraic relation is that 
\bea
\frac{\omega_{M_x,M_y}}{\omega_{M_y,M_x}}=\frac{\omega_{T_1,T_2}}{\omega_{T_2,T_1}}\omega_{(C_4)^2,(C_4)^2}
\eea
and hence $M_xM_yM_x^{-1}M_y^{-1}$ does not give rise to a new invariant. For spinons with $\omega^{e,\epsilon}_{(C_4)^2,(C_4)^2}\equiv1$, we therefore have
\bea
\frac{\omega^{e,\epsilon}_{M_x,M_y}}{\omega^{e,\epsilon}_{M_y,M_x}}=\frac{\omega^{e,\epsilon}_{T_1,T_2}}{\omega^{e,\epsilon}_{T_2,T_1}}
\eea
as reported in Table \ref{table.1}.

\begin{figure}
    \centering
    \includegraphics[width=0.4\columnwidth]{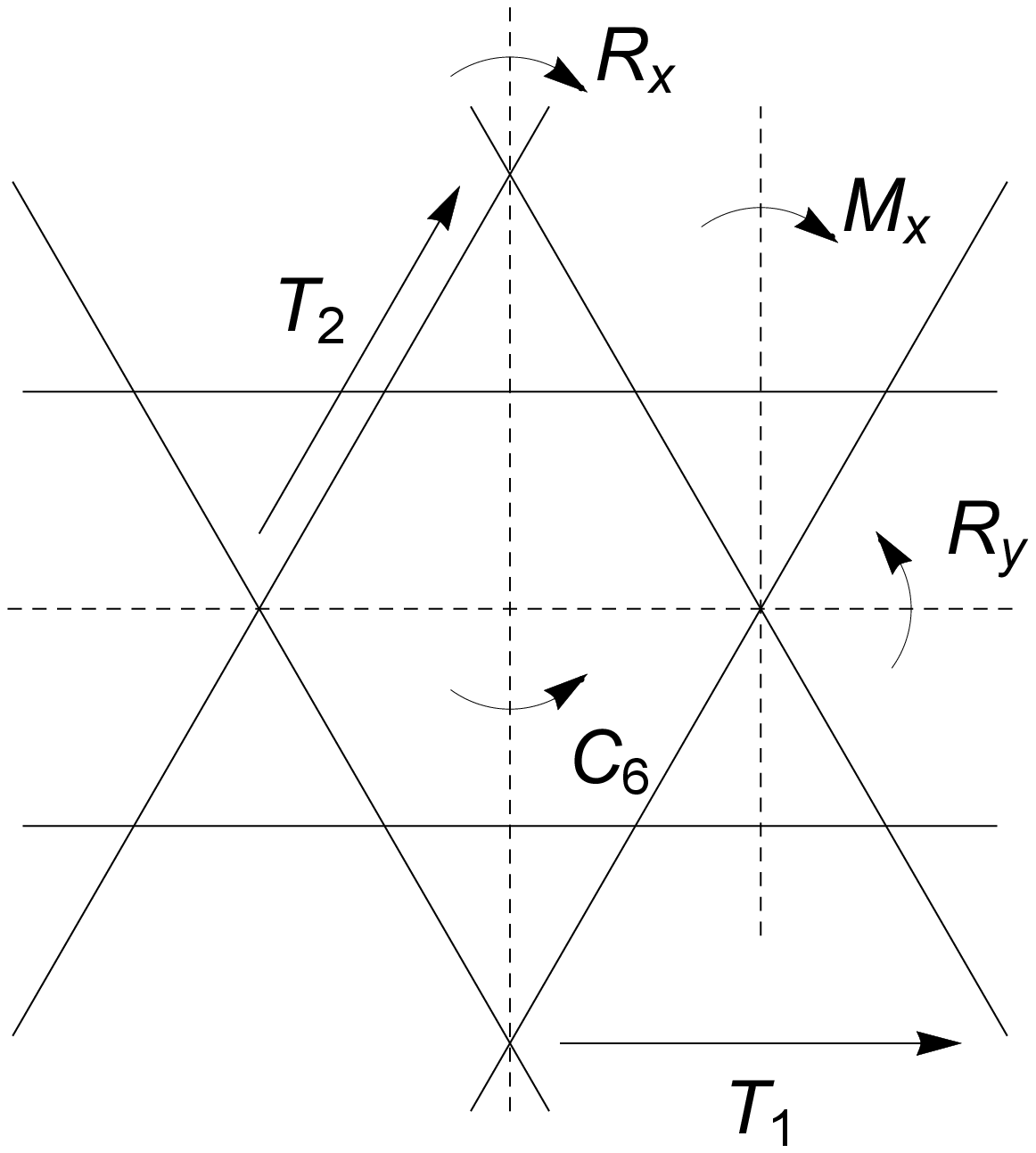}
    \includegraphics[width=0.4\columnwidth]{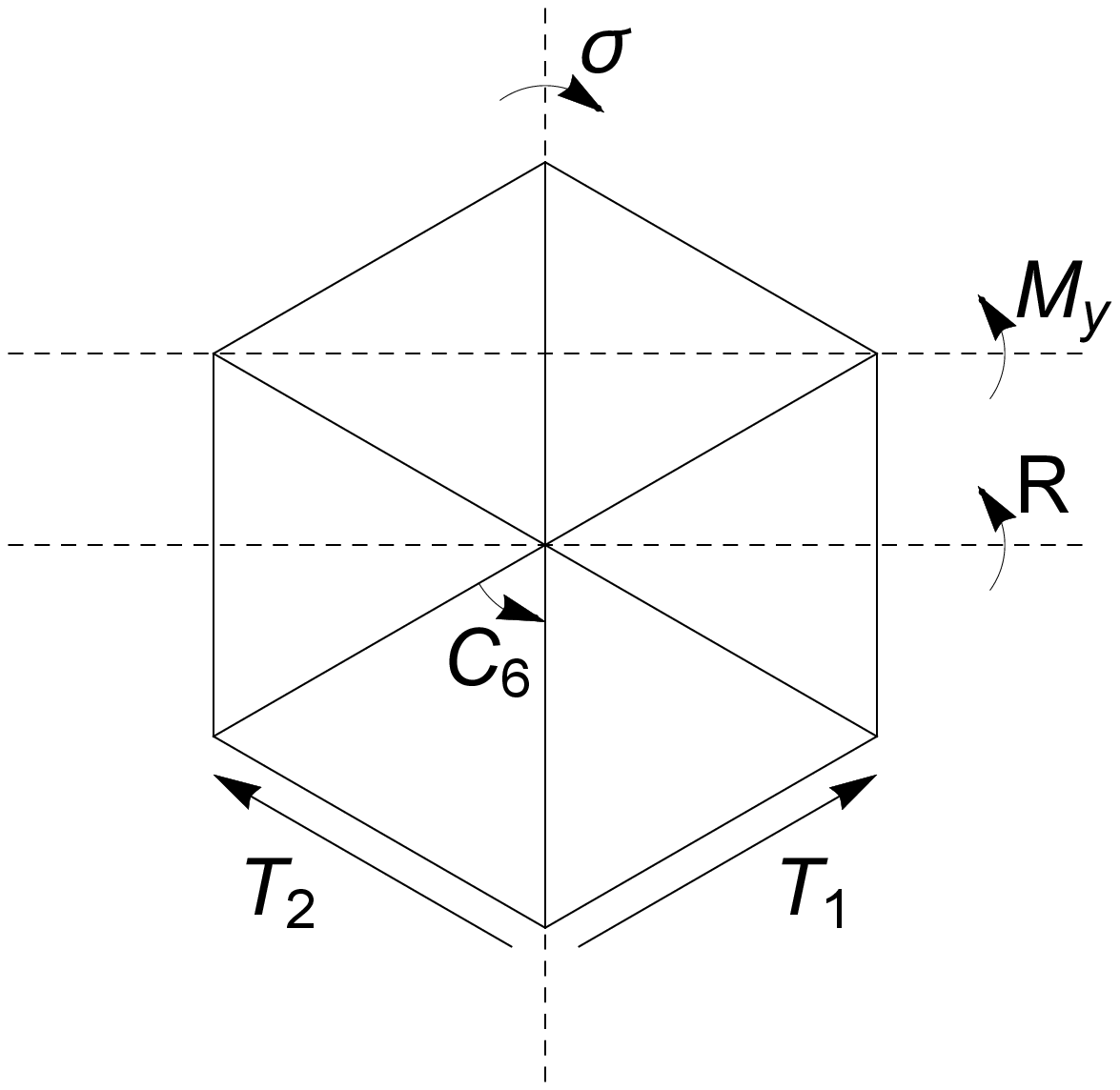}
    \caption{Space group symmetries of the kagome and triangular lattices. }
    \label{fig:kagome+triangle}
\end{figure}

\subsection{Kaogme and triangular lattices}

\begin{table}[h]
    \centering
    \begin{tabular}{|c|c|c|c|c|c|}
    \hline
      Algebraic identities&$[\omega]\in\mathcal{H}^2(G,\mathcal{A})$&$\omega^e$\cite{Wang2006}&$\omega^\epsilon$\cite{Lu2011}&$\omega^m$\cite{Qi2015}&Twist factor $\omega_t$\cite{Qi2015a,Lu_2017}\\
      \hline 
      $\bst^2$&$\omega_{\bst,\bst}$&$-1$&$-1$&1&1\\
      \hline 
      $T_1T_2T_1^{-1}T_2^{-1}$&$\frac{\omega_{T_1,T_2}}{\omega_{T_2,T_1}}$&$(-1)^{p_1}$&$\eta_{12}$&-1&1\\
      \hline
      $(R_y)^2$&$\omega_{R_y,R_y}$&$(-1)^{p_2}$&$\eta_\sigma\eta_{\sigma C_6}$&1&-1\\
    \hline
      $(R_y\bst)^2$&$\omega_{R_y\bst,R_y\bst}$&$-1$&$-\eta_\sigma\eta_{\sigma\bst}\eta_{\sigma C_6}\eta_{C_6\bst}\equiv-1$&1&1\\
    \hline
    $(R_x)^2$&$\omega_{R_x,R_x}$&$(-1)^{p_2+p_3}$&$\eta_\sigma$&1&-1\\
    \hline
    $(R_x\mathcal{T})^2$&$\omega_{R_x\mathcal{T},R_x\mathcal{T}}$&$-1$&$-\eta_\sigma\eta_{\sigma\bst}\equiv-1$&1&1\\
        \hline    
        $M_xR_y M_x^{-1}R_y^{-1}$&$\frac{\omega_{M_x,R_y}}{\omega_{R_y,M_x}}$&$1$&$\eta_{12}\eta_{C_6}\eta_{\sigma C_6}\equiv1$&-1&-1\\
    \hline
    \end{tabular}
    \caption{The classification of gapped $Z_2$ spin liquids of spin-$1/2$'s on the kagome lattice, characterized by 3 independent $Z_2$-valued invariants\cite{Qi2018,Lu_2018}, and their realizations in the Schwinger boson\cite{Wang2006} and Abrikosov fermion\cite{Lu2011} representations.}
    \label{tab:kagome}
\end{table}

Similar to the case of spin-$1/2$'s on the square lattice, one can also classify symmetric $Z_2$ spin liquids of spin-$1/2$'s on the kagome and triangular lattices. Here we briefly review the classification results reported previously in Ref.\cite{Qi2018,Lu_2018}.

In the case of spin-$1/2$'s on the kagome lattice, the space group is $p6mm$, generated by translation $T_1$, hexagon-centered 6-fold rotation $C_6$ and mirror reflection $R_y$. The other symmetries shown in Fig. \ref{fig:kagome+triangle} can be generated as follows:
\bea
&T_2=C_6T_1C_6^{-1},\\
&R_x=(C_6)^3R_y,\\
&M_x=T_1R_x.
\eea
We consider the symmetry group $G=p6mm\times Z_2^\bst$ in addition to $SO(3)$ spin rotational symmetry. The 2nd group cohomology has $\mathcal{H}^2(G,Z_2)=(Z_2)^7$ different classes (see the 7 rows in Table \ref{tab:kagome}), naively leading to $(Z_2)^{14}$ different fractionalization classes. For spin-$1/2$'s on the kagome lattice with the $SO(3)$ spin rotational symmetries, again we can fix many of the cohomology classes and reduce the final classification to only $2^3=8$ distinct symmetric $Z_2$ spin liquids\cite{Lu_2018,Qi2018}. They are characterized by three $Z_2$-valued invariants, as summarized in TABLE \ref{tab:kagome}. 

In particular, since both reflection planes $R_{x,y}$ (see Fig. \ref{fig:kagome+triangle}) cross one site per unit cell, we have 
\bea
\omega^{e,\epsilon}_{R_x\bst,R_x\bst}=\omega^{e,\epsilon}_{R_y\bst,R_y\bst}=-1
\eea
for bosonic spinon $e$ and fermionic spinon $\epsilon$. Meanwhile, since reflection planes $M_x=T_1R_x$ and $R_y$ intersect at one site, we have
\bea
\frac{\omega^{e,\epsilon}_{M_x,R_y}}{\omega^{e,\epsilon}_{R_y,M_x}}=+1
\eea
for both $e$ and $m$. As a result, for the two reflections plane $R_x$ and $R_y$, using the following identity
\bea
\frac{\omega_{M_x,R_y}}{\omega_{R_y,M_x}}\equiv\frac{\omega_{M_x,R_y}}{\omega_{R_y,M_x}}\cdot\frac{\omega_{T_1,T_2}}{\omega_{T_2,T_1}}
\eea
we have
\bea
\frac{\omega^{e,\epsilon}_{M_x,R_y}}{\omega^{e,\epsilon}_{R_y,M_x}}\equiv\frac{\omega^{e,\epsilon}_{T_1,T_2}}{\omega^{e,\epsilon}_{T_2,T_1}}
\eea
as reported in Table \ref{table.2}.\\

\begin{table}[h]
    \centering
    \begin{tabular}{|c|c|c|c|c|c|}
    \hline
      Algebraic identities&$[\omega]\in\mathcal{H}^2(G,\mathcal{A})$&$\omega^e$\cite{Wang2006}&$\omega^\epsilon$\cite{Lu2016}&$\omega^m$\cite{Qi2015}&Twist factor $\omega_t$\cite{Qi2015a,Lu_2017}\\
      \hline 
      $\bst^2$&$\omega_{\bst,\bst}$&$-1$&$-1$&1&1\\
      \hline 
      $T_1T_2T_1^{-1}T_2^{-1}$&$\frac{\omega_{T_1,T_2}}{\omega_{T_2,T_1}}$&$(-1)^{p_1}$&$\eta_{12}$&-1&1\\
      \hline
      $\sigma^2$&$\omega_{\sigma,\sigma}$&$(-1)^{p_2}$&$\eta_\sigma$&1&-1\\
    \hline
      $(\sigma\bst)^2$&$\omega_{\sigma\bst,\sigma\bst}$&$-1$&$-\eta_\sigma\eta_{\sigma\bst}\equiv-1$&1&1\\
    \hline
    $R^2$&$\omega_{R,R}$&$(-1)^{p_2+p_3}$&$\eta_{\sigma C_6}$&1&-1\\
    \hline
    $(R\mathcal{T})^2$&$\omega_{R\mathcal{T},R\mathcal{T}}$&$-1$&$-\eta_{\sigma C_6}\eta_{\sigma\bst}\eta_{C_6\bst}\equiv-1$&1&1\\
        \hline    
        $\sigma R\sigma^{-1}R^{-1}$&$\frac{\omega_{\sigma,R}}{\omega_{R,\sigma}}$&$1$&$\eta_{\sigma}\eta_{C_6}\eta_{\sigma C_6}\equiv1$&-1&-1\\
    \hline
    \end{tabular}
    \caption{The classification of gapped $Z_2$ spin liquids of spin-$1/2$'s on the triangular lattice, characterized by 3 independent $Z_2$-valued invariants\cite{Qi2018,Lu_2018}, and their realizations in the Schwinger boson\cite{Wang2006} and Abrikosov fermion\cite{Lu2016} representations.}
    \label{tab:triangular}
\end{table}

In the case of the triangular lattice, the space group is still $p6mm$ with the same $G=p6mm\times Z_2^\bst$. The spin-$1/2$'s however are located at different Wyckoff sites compared to the kagome lattice. As shown in Fig. \ref{fig:kagome+triangle}, the space group is generated by translation $T_1$, site-centered 6-fold rotation $C_6$ and mirror reflection $R$. The other symmetries in Fig. \ref{fig:kagome+triangle} can be generated as
\bea
&\sigma=(C_6)^3R,\\
&T_2=\sigma T_1\sigma^{-1},\\
&M_y=T_1T_2R.
\eea
Completely in parallel to the kagome lattice case, one can show the $\mathcal{H}^2(G,\mathcal{A})=(Z_2)^{14}$ cohomology classes summarized in Table \ref{tab:triangular} can be reduced to $2^3$ distinct symmetric $Z_2$ spin liquids, by requiring a spectrum gap and $SO(3)$ spin rotational symmetry\cite{Qi2018,Lu_2018}. They are listed in Table \ref{table.3}. In particular, following the group cohomology identity:
\bea
\frac{\omega_{\sigma,M_y}}{\omega_{M_y,\sigma}}\equiv\frac{\omega_{\sigma,R}}{\omega_{R,\sigma}}\cdot\frac{\omega_{T_1,T_2}}{\omega_{T_2,T_1}}
\eea
we have
\bea
\frac{\omega^{e,\epsilon}_{\sigma,M_y}}{\omega^{e,\epsilon}_{M_y,\sigma}}\equiv\frac{\omega^{e,\epsilon}_{T_1,T_2}}{\omega^{e,\epsilon}_{T_2,T_1}}
\eea
as reported in Table \ref{table.3}.

\section{Exactly solvable model}\label{section.1}
We use an exactly solvable model with spin-$1/2$ spinon excitations to illustrate the symmetry protected non-Kramers doublet localized at the impurity, in the Kondo screening phase. The model is based on Kitaev's toric code on the square lattice, with four types of topological excitations, including the trivial sector $1$, bosonic spinon $e$, vison $m$, and fermionic spinon $\epsilon=e\times m$. To incorporate the $SU(2)$ symmetry, we revise the toric code to construct a $Z_2$ spin liquid where $e,\epsilon$ excitations each carry spin-$1/2$.

\begin{figure}[h]
    \centering
    \includegraphics[width=0.48\columnwidth]{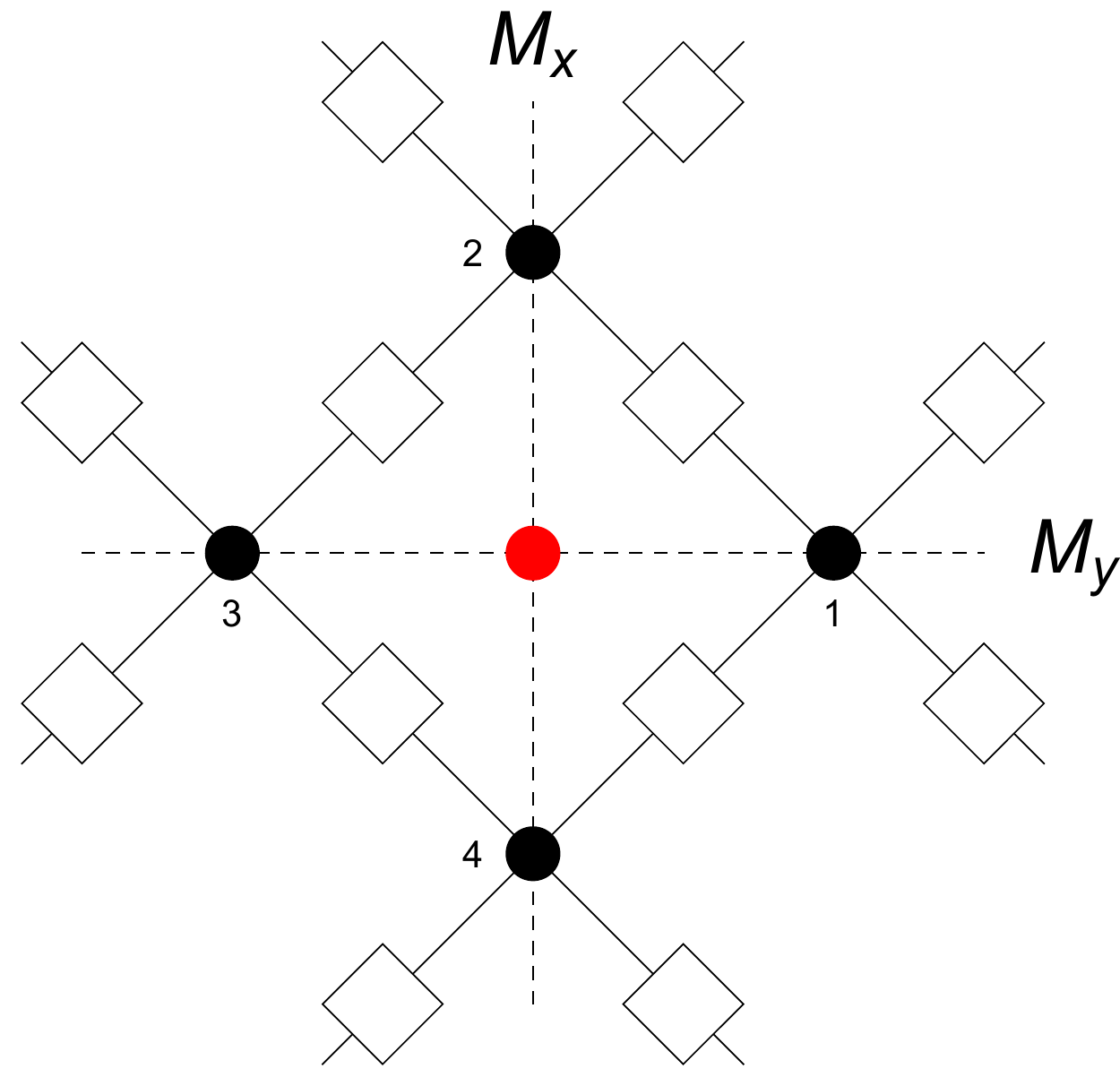}
    \caption{An exactly solvable model for a spin-$1/2$ impurity in a $Z_2$ spin liquid with $SU(2)$ symmetry. The squares on link centers are qubits in the toric code\cite{Kitaev_2003}. Each black dot on a star/vertex represents a Hilbert space of spin $0\oplus 1/2$. The red dot denotes the spin-$1/2$ impurity with a site symmetry $G_s=Z_2^{M_x}\times Z_2^{M_y}$, which is coupled to four neighboring spin-$1/2$'s.}
    \label{fig:5}
\end{figure}

The Hilbert space of our model consists of two parts in the bulk, as shown in Fig. \ref{fig:5}. Each squares on link center represents a qubit interacting with each other as in the toric code\cite{Kitaev_2003}. Each black dots on the vertices stands for a three dimensional linear space constructed as spin $0 \oplus 1/2$. This Hilbert space allows us to attach a spin $1/2$ to each $e$ particle on the vertex. In the ground state, every vertex spin is in the spin-$0$ state, but if there is a $e$ excitation on a vertex, the vertex state is switched into spin-$1/2$. The red dot represents a spin-$1/2$ impurity interacting with four nearest spin-$1/2$'s. 
 
To be more specific, the model is
\begin{equation}
    \hat{H}_{bulk}= -\sum_s A_s -\sum_p B_p -\sum_s \Delta\big(\frac{(A_s+1)}2 P_s(S=0)+\frac{(1-A_s)}2 P_s(S=1/2)\big)\label{eq:bulk model}
\end{equation}
where $s$ means star (or vertex) and $p$ means plaquette on the square lattice of solid lines. $A_s$ and $B_p$ are corresponding star term and plaquette term in the toric code\cite{Kitaev_2003}: 
\begin{equation}\label{def:vertex+plaquette}
    A_s= \prod_{j\in s} \sigma_j^x,\quad B_p=\prod_{j\in p} \sigma_j^z
\end{equation}
The second part of Hamiltonian describes the coupling between spins and the toric code model. We assume $\Delta \gg 1$. $P_s(S=0)$ means projection operator on vertex $s$ into $S=0$ subspace if $A_s=1$ or no $e$ excitation. Otherwise, if $A_s=-1$, the $S=1/2$ subspace has a lower energy because of the $\Delta$ term. Note the difference between capital $P,S$ and $p,s$. They have very different meanings here.

The $SU(2)$ symmetry is implemented in the Hilbert space of spin $0\oplus 1/2$ on each vertex. The model shares the same ground state as toric code where all vertex spins are in the spin-$0$ state. If we choose a very large $\Delta\gg1$, the low energy excitations are spin-0 $m$ particles, and $e,\epsilon$ particles each carry spin-$1/2$. This model preserves two mirror symmetries $M_x, M_y$ shown as dotted line in Fig. \ref{fig:5}. The symmetry fractionalization class regarding the site symmetry $G_s=Z_2^{M_x}\times Z_2^{M_y}$ is given by\cite{Essin2013} 
\bea\label{eq:toric code frac class}
\frac{\omega^e_{M_x,M_y}}{\omega^e_{M_y,M_x}}=\frac{\omega^m_{M_x,M_y}}{\omega^m_{M_y,M_x}}=+1,~~~\frac{\omega^\epsilon_{M_x,M_y}}{\omega^\epsilon_{M_y,M_x}}=-1.
\eea
In other words, only the fermionic spinon $\epsilon$ has a nontrivial fractionalization class associated with two mirror symmetries $M_{x,y}$. 

The coupling between impurity and bulk also preserves $SU(2)$ symmetry. To start, we couple the spin-$1/2$ impurity to the four nearly spin-$1/2$ degrees of freedom in the bulk with antiferromagnetic Heisenberg terms: 
\begin{equation}
    H_{imp} = J \sum_{\langle i, imp \rangle} \vec{S}_i \cdot \vec{S}_{imp}
\end{equation}
where $i$ is summed over four spins closest to the impurity. When $J$ is positive, impurity couples to bulk spins antiferromagnetically. The full Hamiltonian is 
\begin{equation}
    \hat{H}=\hat{H}_{bulk}+\hat{H}_{imp}
\end{equation}

Now we study the ground state with an impurity. When $J$ is small, the impurity spin remains an unscreened spin-$1/2$, and the decoupled bulk stays in the original ground state as the toric code. This is the unscreened phase. When $J> 8/9$, the system enters into a new phase where $H_{imp}$ is minimized. Since the impurity attracts four spin-$1/2$ $e$ particles around it, this is an overscreened phase. 
\begin{equation}
    |G_\text{overscreened}\rangle = | e_1,e_2,e_3,e_4\rangle \otimes | S_{tot}=3/2, S_4=2,S_{imp}=1/2\rangle
\end{equation}
Here, $e_i, i=1\cdots 4$ means four e particles on vertices $1\cdots 4$. There are two good quantum numbers to label the ground state: $S_4$ denotes the total spin of four spin $1/2$, $S_{tot}$ means the total spin of the whole system. 

In order to achieve a Kondo screening phase with a paramagnetic ground state, we add another coupling to the impurity Hamiltonian:
\begin{equation}\label{eq:impurity term}
    H_{imp}^\prime= H_{imp} + E_c(A_1+A_2+A_3+A_4-3)^2
\end{equation}
where $\hat A_i$ denotes the vertex term (\ref{def:vertex+plaquette}) on vertex $i$. The second term can be viewed as a Coulomb repulsion of strength $E_c>0$ for spinons near the impurity site, since $(A_1+A_2+A_3+A_4-3)^2$ is minimized when there is no $e$ particles or only one $e$ particle. Let's assume $E \gg 1,J$. When $J> 4/3$, in the ground state the impurity spin-$1/2$ favors to form a singlet bound state with one $e$ particle from the bulk, which also brings a spin-$1/2$ at one neighboring site of the impurity. 
\begin{equation}
    |G_i\rangle = | e_i\rangle \otimes (|\uparrow_i \downarrow_{imp}\rangle - |\downarrow_i \uparrow_{imp}\rangle)
\end{equation}
Here $i$ can be $1\cdots 4$. $|\uparrow_i\rangle$ means the vertex $i$ is in the spin-up state in the 3-dimensional space. The ground state is four fold degenerate. In particular, if we choose the following gauge
\bea
|e_2\rangle=\sigma^z_{12}|e_1\rangle,~~~|e_3\rangle=\sigma^z_{12}\sigma^z_{23}|e_1\rangle,~~~|e_4\rangle=\sigma^z_{14}|e_1\rangle
\eea
the four degenerate states transform under crystal symmetries as follows:
\bea\label{eq:mx}
&|G_i\rangle\overset{M_x}\longrightarrow \sum_j U(M_x)_{i,j}|G_j\rangle,~~~U({M_x})=\begin{pmatrix}0&0&1&0\\0&1&0&0\\1&0&0&0\\0&0&0&B_p[1234]\end{pmatrix}=U^{-1}(C_4)U(M_y)U(C_4),\\
\label{eq:my}&|G_i\rangle\overset{M_y}\longrightarrow \sum_j U(M_y)_{i,j}|G_j\rangle,~~~U({M_y})=\begin{pmatrix}1&0&0&0\\0&0&0&1\\0&0&B_p[1234]&0\\0&1&0&0\end{pmatrix},\\
\label{eq:c4}&|G_i\rangle\overset{C_4}\longrightarrow \sum_j U(C_4)_{i,j}|G_j\rangle,~~~U({C_4})=\begin{pmatrix}0&1&0&0\\0&0&1&0\\0&0&0&B_p[1234]\\1&0&0&0\end{pmatrix}
\eea
where $C_4$ is the 4-fold rotation around the plaquette center (i.e. the impurity site). We have defined the $Z_2$-valued variable: 
\bea
B_p[1234]\equiv\langle\hat B_p[1234]\rangle=\langle0|\sigma^z_{12}\sigma^z_{23}\sigma^z_{34}\sigma^z_{41}|0\rangle=\pm1
\eea
where $|0\rangle$ is the ground state of the bulk model (\ref{eq:bulk model}), and $\hat B_p[1234]$ is the plaquette term for the plaquette including sites $1,2,3,4$. It is instructive to work in the basis of $C_4 $eigenstates and we shall follow this strategy in the following calculations. 

In the case of $B_p[1234]=+1$, the $C_4$ eigenstate $|n\rangle$ with eigenvalue $e^{in\pi/2}$ is given by
\bea
|n=0,1,2,3\rangle=\frac12\sum_{j=1}^4 e^{in(j-1)\pi/2}|G_j\rangle
\eea
In this case both $|n=0\rangle$ and $|n=2\rangle$ states are invariant under $M_{x,y}$ operations and hence can be the unique ground state of the system. For example, to select $|n=0\rangle=\sum_j|G_j\rangle/4$ as the unique ground state, we can add the following perturbation
\bea\notag
&\Delta H_{imp}=-V\sum_i|G_i\rangle\langle G_{i+1}|+h.c.\\
&=-V\sum_{i=1}^4\sigma^z_{i,i+1}\otimes\big[(|\uparrow_i \downarrow_{imp} 0_{i+1}\rangle-|\downarrow_i \uparrow_{imp} 0_{i+1}\rangle) (\langle\uparrow_{i+1} \downarrow_{imp} 0_{i}|-\langle\downarrow_{i+1}\uparrow_{imp} 0_i|)+h.c.\big]
\eea
which is symmetric w.r.t. all site symmetries at the impurity site. In the presence of this term, in the Kondo screened phase, the system will have a unique paramagnetic ground state with no protected degeneracy at the impurity site. 

In the case of $B_p[1234]=-1$ which implies one $m$ particle present in the plaquette, on the other hand, the anyon screening the spin-$1/2$ impurity is not $e$, but instead a fermionic spinon $\epsilon=e\times m$. This can be achieved by adding one extra term to (\ref{eq:impurity term})
\bea
H^{\prime\prime}_{imp}=H^\prime_{imp}+\Delta_\epsilon\hat B_p[1234]
\eea
When $\Delta_\epsilon>1$, the ground state of the impurity Hamiltonian would favor the low energy fermionic spinon $\epsilon$ over bosonic spinon $e$, and therefore screen the impurity spin-$1/2$ by one $\epsilon$ particle. We can similarly label the 4 ground states in the low-energy manifold of the full Hamiltonian $H_{bulk}+H^{\prime\prime}_{imp}$ as
\begin{equation}
    |G_i\rangle = | \epsilon_i\rangle \otimes (|\uparrow_i \downarrow_{imp}\rangle - |\downarrow_i \uparrow_{imp}\rangle)
\end{equation}
Choosing the same gauge, their symmetry transformation laws still follow Eqs. (\ref{eq:mx})-(\ref{eq:c4}). If we label them in the basis of $C_4$ eigenstates
\bea
&C_4|\pm1\rangle=e^{\pm i\pi/4}|\pm1\rangle,~~~|\pm1\rangle\equiv\frac{e^{\pm i\pi/4}|G_2\rangle\pm i|G_2\rangle+e^{\pm i3\pi/4}|G_3\rangle+|G_4\rangle}2;\\
&C_4|\pm3\rangle=e^{\pm i3\pi/4}|\pm3\rangle,~~~|\pm3\rangle\equiv\frac{e^{\pm i3\pi/4}|G_2\rangle\mp i|G_2\rangle+e^{\pm i\pi/4}|G_3\rangle+|G_4\rangle}2.
\eea
Each of the two doublets, i.e. $|\pm1\rangle$ (or $|\pm3\rangle$) forms a 2-dimensional irreducible projective representation of the site symmetry group $G_s=Z_2^{M_x}\times Z_2^{M_y}$. In particular, in the doublet $|\pm1\rangle$ with $C_4$ eigenvalues $e^{\pm i\pi/4}$, the two mirror actions are represented by Pauli matrices:
\bea
\langle\pm1|U(M_x)|\pm1\rangle=-\sigma_x,~~\langle\pm1|U(M_y)|\pm1\rangle=\sigma_y.
\eea
Similarly, in the doublet $|\pm3\rangle$ with $C_4$ eigenvalues $e^{\pm i3\pi/4}$, the two mirror actions are represented by Pauli matrices:
\bea
\langle\pm3|U(M_x)|\pm3\rangle=-\sigma_x,~~\langle\pm3|U(M_y)|\pm3\rangle=-\sigma_y.
\eea
Indeed $M_x$ and $M_y$ operations anticommute with each other in each doublet pair, consistent with the fractionalization class (\ref{eq:toric code frac class}) of fermionic spinon $\epsilon$ in the toric code. As a result, any symmetry-preserving local Hamiltonian cannot lift this two-fold degeneracy, which is protected by two mirror symmetries $M_{x,y}$.

\section{Large-$N$ theory and thermodynamics}\label{section.2}
To further illustrate the effect of nontrivial symmetry fractionalization classes on Kondo impurity problem in $Z_2$ spin liquids, from the perspective of parton construction\cite{Lu_2018}, we construct a large-$N$ model for the Kondo impurity problem in a symmetric $Z_2$ spin liquid. We then carry out large-$N$ mean-field calculations to obtain the phase diagram, and the temperature dependence of thermodynamic quantities including the specific heat (and hence entropy) and uniform susceptibility. 

We start from a solvable model similar to Kitaev's honeycomb model \cite{Kitaev2006} for the bulk $Z_2$ spin liquid, where the exact spectrum of the system is given by parton mean-field ansatz under different $Z_2$ flux configurations. In these models, the fractionalization class $[\omega^\epsilon]\in\mathcal{H}^2(G,z_2)$ of fermionic spinons $\epsilon$ manifest itself as a projective symmetry group (PSG)\cite{Wen2002} of the symmetry group $G$ carried by fermionic partons. This model is defined on the square lattice and can support gapped $Z_2$ spin liquid ground states with either nontrivial or trivial PSGs of fermionic partons (or spinons). The model is not exactly solvable anymore after the impurity spin is added and coupled to the bulk, but we can still obtain analytical results in large-$N$ limit.

\subsection{Single impurity in the $SU(2)$ parton construction}

\begin{figure}[h]
    \centering
    \includegraphics[width=0.4\columnwidth]{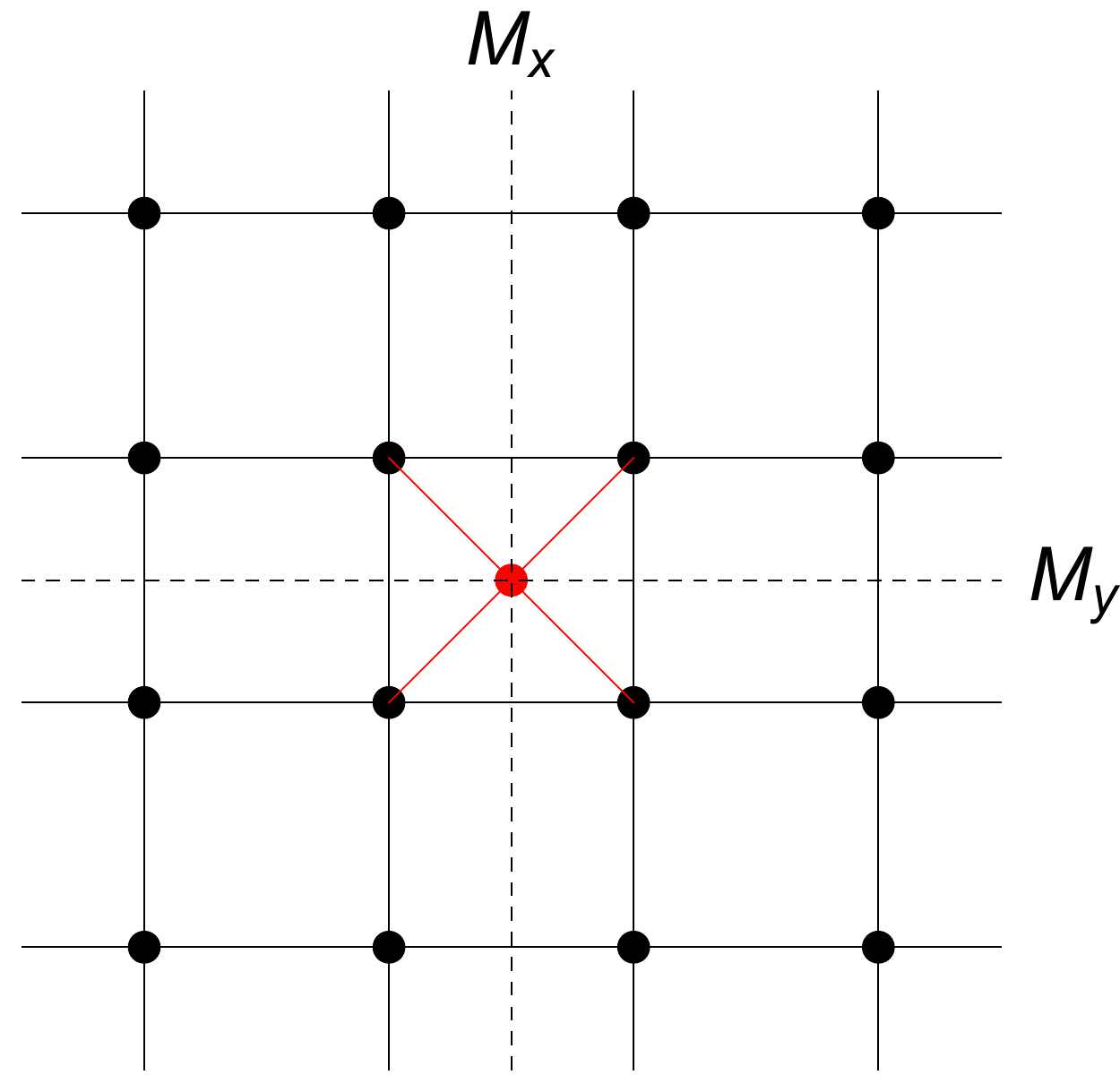}
    \includegraphics[width=0.5\columnwidth]{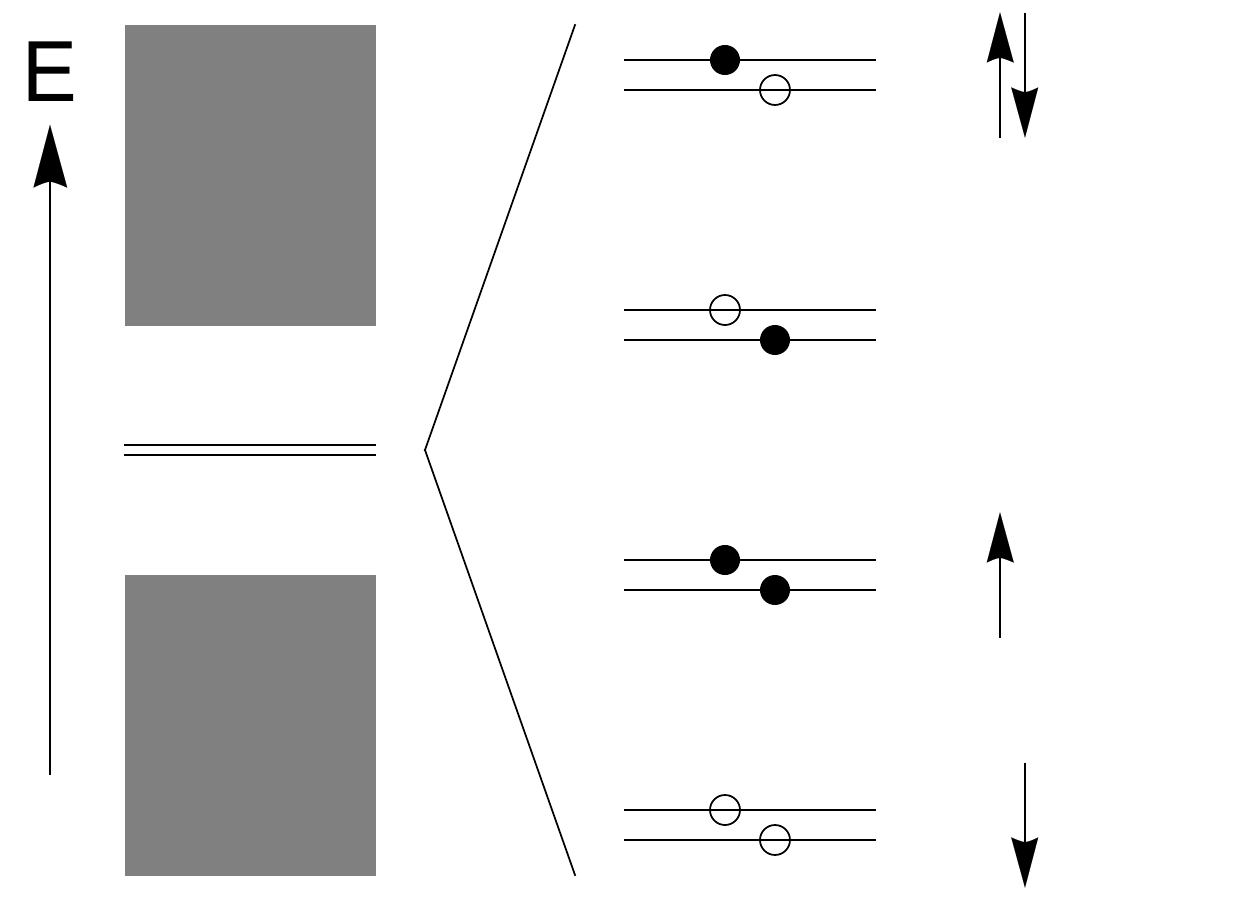}
    \caption{ (Left) A spin-$1/2$ impurity at a plaquette center on a square lattice, preserving two mirror symmetries $M_{x,y}$. (Right) The spectrum of the parton BdG Hamiltonian with a nontrivial PSG (\ref{eq.2}), where two mirror symmetries enforce a pair of zero modes on an even by even square lattice. Among the 4 many-body states obtained by filling the two zero modes, only the both filled and the both empty states are physical, corresponding to the unscreened phase with a free spin at the impurity site.}
    \label{fig:7}
\end{figure}

The parton construction of quantum spin liquids\cite{Wen2002} provides another aspect to look at the Kondo effect in $Z_2$ spin liquids. We consider a square lattice with a spin-$1/2$ on each site. In the $SU(2)$ parton (or slave particle) construction, spin $1/2$ can be represented by a doublet of fermions $\{c_{\uparrow},c_{\downarrow}\}$ on each site with the single occupancy constraint $c^\dagger_\uparrow c_\uparrow+c^\dagger_\downarrow c_\downarrow =1$\cite{Wen2002}. In this way, the physical spin state can be obtained by projecting a many-fermion state into the physical spin Hilbert space, by enforcing the single-occupancy constraint on each site. Below, we use the $SU(2)$ parton construction on the square lattice, as shown in Fig.\ref{fig:7}, to clarify the physical reason behind the symmetry protected non-Kramers doublet localized at the impurity site, in the Kondo screening phase. 

In the parton construction, in order to obtain a symmetric spin state, the fermion state only needs to be invariant under symmetry up to a gauge transformation. In other words, the fermions preserve a projetive symmetry group (PGS), which is an extension of the physical symmetry group by an invariant gauge group (IGG)\cite{Wen2002}. The IGG is $Z_2$ in our case of $Z_2$ spin liquids. Specifically, we consider two mirror symmetries $M_x,M_y$, which always commute with each other in the physical spin system. However for fermionic spinons with a nontrivial fractionalization class, they satisfy the following PSG:
\begin{equation}\label{eq.2}
    M_x M_y M_x^{-1} M_y^{-1} = (-)^{\hat{N}_F}
\end{equation}
where  $\hat{N}_F$ is total fermion number operator. In contrast, for a $Z_2$ spin liquid with a trivial fractionalization class, the fermionic spinons satisfy $M_x M_y M_x^{-1} M_y^{-1} = 1$. 

Due to $SU(2)$ spin rotational symmetry, a quadratic mean-field ansatz for a symmetric $Z_2$ spin liquid has the following form\cite{Wen2002}
\begin{equation}\label{Eq.bulk}
    H= \sum_{i,j} J_{i,j}\psi_i^\dagger u_{i,j} \psi_j+h.c. 
\end{equation}
where we defined 2-component fermions $\psi_i= (c_\uparrow,c^\dagger_\downarrow)^T$. The particle-hole pair $\gamma_{\pm E}$ with energy $E$ and $-E$ always appears together in the spectrum of the parton BdG Hamiltonian in the Nambu basis, as guaranteed by time reversal symmetry $\mathcal{T}$ or spin rotational symmetry $e^{i\pi S^y}$. Both $\mathcal{T}$ and $e^{i\pi S_y}$ symmetries map $\psi$ to $(c_\downarrow,-c^\dagger_\uparrow)^T=i\tau^y\psi_i^\ast$, therefore mapping a mode $\gamma_E$ to $\gamma_{-E}^\dagger$.

Now we place a spin-$1/2$ impurity to one plaquette center (see Fig.\ref{fig:7}) and couple it to neighboring spins. Similar to bulk spin-$1/2$'s, the impurity spin-$1/2$ can also be represented by a doublet of fermions. We consider a $ N\times N$ square lattice with $N=$~even, and one impurity at the center the torus. Together with the impurity site, overall there is an odd number of spin-$1/2$s, and $2(N^2+1)$ energy levels in the spectrum of the BdG Hamiltonian. 

In the case of a nontrivial fractionalization class with $M_x M_y M_x^{-1} M_y^{-1}=(-1)^{\hat N_F}$, each energy level in the spectrum of the parton BdG Hamiltonian must be at least 2-fold degenerate, forming a two dimensional projective representation of mirror symmetries $Z_2^{M_x}\times Z_2^{M_y}$. Here we discuss two different scenarios, corresponding to the unscreened phase and the Kondo screening phase in the parton construction. 

First, if the mean-field ansatz (including impurity) preserves mirror symmetries $M_{x,y}$ mentioned above, there must be a pair of zero modes in the parton BdG spectrum. This is because the each energy level must be 2-fold degenerate due to PSG (\ref{eq.2}), and therefore any particle-hole symmetric spectrum must have $N^2$ negative energy levels, $N^2$ positive energy levels, and 2 degenerate zero modes, as shown in Fig.\ref{fig:7}.  
Naively there are four degenerate many-body ground states, because those two zero modes can be either filled or empty, as shown in Fig. \ref{fig:7}. Both filled or both empty in the Nambu basis corresponds to unscreened free spin $\uparrow$ or $\downarrow$ state, while filling only one zero mode leads to paramagnetic ground states. Here due to the single occupancy constraint on each site, the physical Hilbert space must have an odd number of fermions on a lattice with an odd number of sites, and hence an even particle number in the Nambu basis. As a result, only the both filled and the both empty states are physical, and adiabatically connected to the free spin state where partons on the impurity site do not couple to bulk partons in the BdG ansatz. This corresponds to an unscreened phase with a free spin on the impurity site. 

The second scenario is when the parton BdG spectrum is gapped because the mean-field ansatz around the impurity site spontaneously breaks the mirror symmetries, even though the bulk spin liquid preserves both mirrors. In this case, due to lack of mirror symmetries in the full BdG ansatz, there should be $N^2+1$ negative levels, and $N^2+1$ positive ones, and the BdG spectrum is gapped with a unique many-body ground state. This corresponds to the Kondo screening phase with paramagnetic ground states. Here we need to consider the two symmetry-breaking mean-field ansatz $|A_1\rangle$ and $|A_2\rangle$ related by mirror operation $M_x$ (or $M_y$). As illustrated in Fig. \ref{fig:8}. One can show that due to nontrivial PSG (\ref{eq.2}), quantum tunneling between these two states are forbidden by the mirror symmetries, leading to a 2-fold degeneracy that cannot be lifted by any symmetry-preserving local Hamiltonian. This exactly corresponds to the mirror symmetry protected non-Kramers doublet at the impurity site, in the Kondo screening phase. More details will be discussed later in the self-consistent calculations of the $Sp(2N)$ model. 

For $Z_2$ spin liquid with a trivial spinon PSG, i.e. when $M_x M_y M_x^{-1} M_y^{-1}=1$ for partons, the symmetry fractionalization of $\epsilon$ particle is trivial. In this case, the BdG spectrum is generically gapped with $N^2+1$ positive and negative levels. The unique ground state preserves both mirror symmetries and is a paramagnet, corresponding to a usual Kondo screening phase, with no symmetry protected degeneracy. This physical picture based on $SU(2)$ parton construction will be manifested in more detail soon in the large-$N$ $Sp(2N)$ model.

Up to this point, we have only discussed a single spin-$1/2$ impurity on a lattice with an even number of spin-$1/2$s. In this situation, in fact, the Kondo screening phase will also excite another delocalized spinon in the bulk, when one spinon is localized at the impurity site to form a singlet bound state with the impurity. In a realistic finite size calculation, to achieve a paramagnetic ground state, we always consider two high-symmetry impurity sites, separated from each other by a largest distance. In this case, there will be one non-Kramers doublet at each impurity site, in the Kondo screening phase.

\subsection{The $Sp(2N)$ model}
The full Hamiltonian of the large $N$ theory contains the bulk $Z_2$ spin liquid and an impurity coupled to bulk.  
\begin{equation}
    H=H_{bulk}+H_{imp}
\end{equation}

\begin{figure}[h]
    \centering
    \includegraphics[width=0.5\columnwidth]{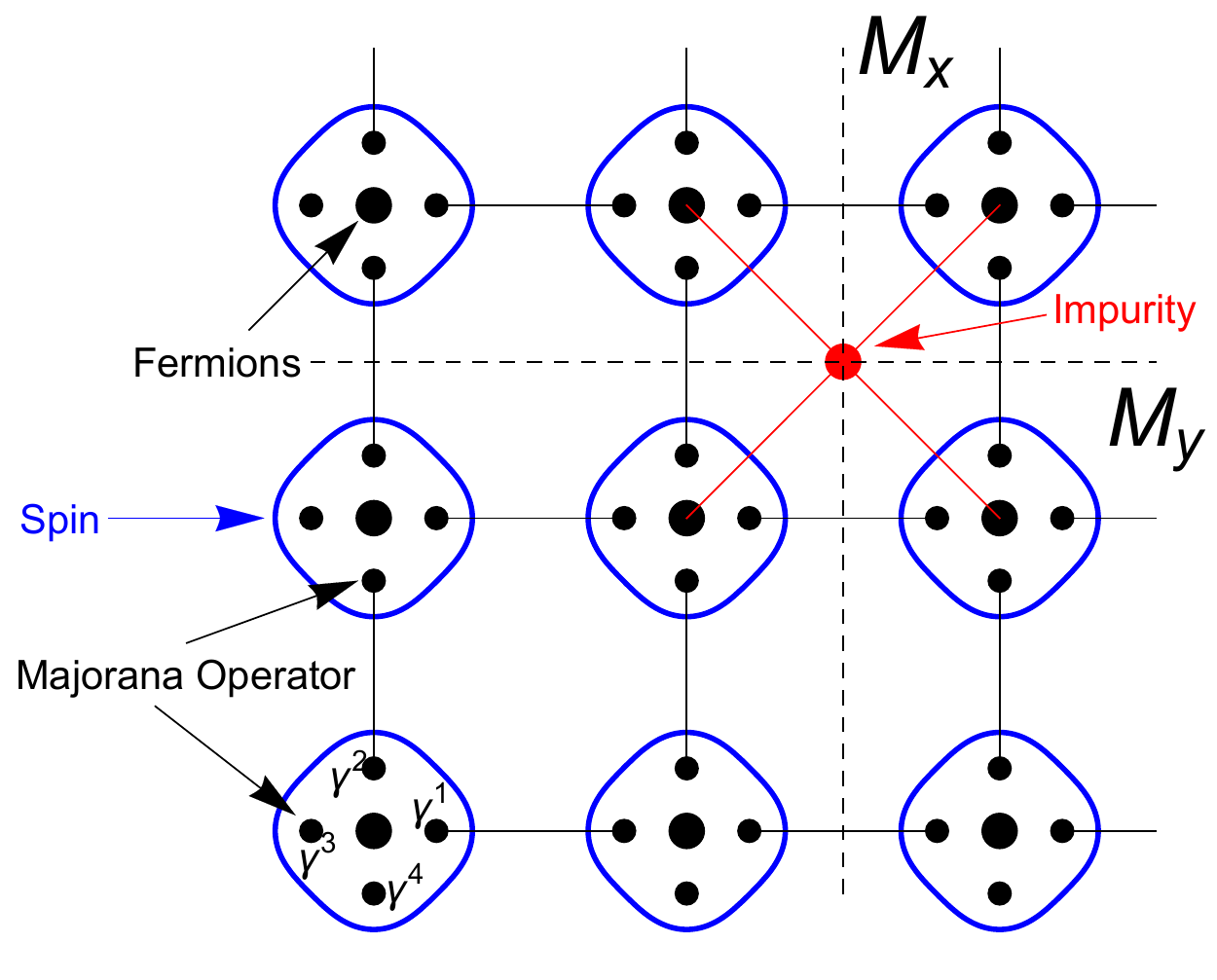}
    \caption{An illustration of the solvable $Sp(2N)$ model of a $Z_2$ spin liquid. Each blue circle represents a physical spin composed of four Majorana operators (small black dots) and $2N$ complex fermions (big black dot). They live on a square lattice, with the two mirror symmetries shown by dashed lines. Red dot denotes the impurity and it only couples to the four nearest spins.}
    \label{fig.model}
\end{figure}

As shown in Fig. \ref{fig.model}, on each site of the two-dimensional square lattice, we define a physical ``spin'' (more precisely, a qudit) composed of $2N$ fermions and $4$ Majorana fermions with a total fermion parity constraint shown below. We have $4$ Majorana operators $\gamma^\alpha_i$, where $i$ labels lattice sites and $\alpha=1\cdots 4$ labels four species of Majorana fermions on the same site. We also have $N$ flavors of spin-$1/2$ complex fermions $c_{i\alpha}^{a}$, where $i$ labels lattice sites, $\alpha=\uparrow,\downarrow$ labels spin and $a=1\cdots N$ labels flavor. The aforementioned constraint is that the total fermion parity on each site is even:
\begin{equation}
    \prod_{\alpha=1}^4 \gamma^\alpha (-1)^{\sum_{a=1}^N (f_{i\uparrow}^\dagger f_{i\uparrow}+f_{i\downarrow}^\dagger f_{i\downarrow})} =1
\end{equation}

The solvable model is illustrated in Fig. \ref{fig.model}. On each site blue circle represents spin, four small dots represent four Majorana operators and the big dot at the center represents $2N$ fermions. Coupling between Majorana operators happens between neighbour sites as shown by solid lines in the Figure. We label two Majorana fermions which forms a dimer between neighboring sites $i,j$ as $\gamma_i^{\alpha(i,j)}$ and $\gamma_j^{\beta(i,j)}$, where we defined 
\begin{equation}
\begin{aligned}
\alpha(i,i+\hat{x})&=1 \quad \beta(i,i+\hat{x})&=3\\ 
\alpha(i,i+\hat{y})&=2 \quad \beta(i,i+\hat{y})&=4\\
\alpha(i,i-\hat{x})&=3 \quad \beta(i,i-\hat{x})&=1\\
\alpha(i,i-\hat{y})&=4 \quad \beta(i,i-\hat{y})&=2
\end{aligned}
\end{equation}

The solvable Hamiltonian for the bulk $Z_2$ spin liquid is 
\begin{equation}
\begin{aligned}
H_{bulk}=t&\sum_{\langle i,j\rangle} \hat{u}_{i,j} (ic_{i\uparrow}^{a\dagger} c^a_{j\uparrow}+ic_{i\downarrow}^{a\dagger} c^a_{j\downarrow}+h.c.)\\
+t^\prime&\sum_{j=i\pm 2\hat{x}+2\hat{y}} \hat{u}_{i,j}(c_{i\uparrow}^{a\dagger} c^a_{j\uparrow}+c_{i\downarrow}^{a\dagger} c^a_{j\downarrow}+h.c.)\\
+&\Delta \sum_{i} c^{a}_{i\uparrow}c^a_{i\downarrow}+g \sum_i \hat{u}_{i,i+\hat{x}}\hat{u}_{i+\hat{x},i+\hat{x}+\hat{y}}\hat{u}_{i+\hat{x}+\hat{y},i+\hat{y}}\hat{u}_{i+\hat{y},i}
\end{aligned}
\end{equation}
The first term is the nearest neighbour coupling of physical spins, where $\hat{u}_{i,j}$ is a conserved quantity (vector potential of the emergent $Z_2$ gauge field) of this Hamiltonian, defined as\cite{Kitaev2006} 
\begin{equation}
    u_{i,j}=i\gamma^{\alpha(i,j)}_i\gamma^{\beta(i,j)}_j
\end{equation}
Here $\alpha(i,j)$ and $\beta(i,j)$ are two functions of two nearby sites $i,j$ defined previously. The second term with strength $t^\prime$ is interaction between $4$th nearest neighbours. To construct this interaction terms we also use $\hat{u}_{i,j}$ for two sites $i,j$ that are not nearest neighbours. We construct it by multiplying a series of connected nearest neighbor $u_{\langle i,j\rangle}$s. For example, we define
\begin{equation}
    u_{i,i+2\hat{x}+2\hat{y}}=u_{i,i+\hat{x}} u_{i+\hat{x},i+2\hat{x}} u_{i+2\hat{x},i+2\hat{x}+\hat{y}} u_{i+2\hat{x}+\hat{y},i+2\hat{x}+2\hat{y}}
\end{equation}
We can also change all $+$ sign to $-$ sign before $\hat{x}$ or $\hat{y}$ in the equation above to define $u_{i,i-2\hat{x}+2\hat{y}}$, $u_{i,i+2\hat{x}-2\hat{y}}$ and $u_{i,i-2\hat{x}-2\hat{y}}$.
The third term represents for the onsite pairing. The fourth term with a positive $g>0$ selects the flux around each plaquette in the ground state. It is assumed to be much larger than the other 3 terms to make sure that $\pi$ flux has the lowest energy.

This Hamiltonian preserves time reversal symmetry $\mathcal{T}$. Time reversal symmetry does not change $\gamma_i^\alpha$ and it is represented as $\mathcal{T}=i\sigma^2 K$  on the fermions, where $\sigma^2$ is Pauli matrix of spin index and $K$ is the complex conjugation operation. Translational symmetries $T_x,T_y$ and link centered mirror symmetry $M_x, M_y$ (labeled in Fig. \ref{fig.model}) act conventionally on the lattice site indices and do not act on the spin indices.

Since the link variables $\{\hat{u}_{i,j}\}$ commute with each other and Hamiltonian, they can be diagonalized simultanenously and fixed to $u_{i,j}=\pm1$. In the absence of vison excitations, in the ground state with a $\pi$ flux per plaquette, the effective Hamiltonian of the bulk for the complex fermions is described by 
\begin{equation}\label{Eq.bulk}
    H_{bulk} = \sum_{i,j} J_{i,j}\psi_i^\dagger u_{i,j} \psi_j+h.c. 
\end{equation}
where 
\begin{equation}\label{eq.4}
\begin{aligned}
    &u_{i,i+\hat{x}}=it\tau^0\\
    &u_{i,i+\hat{y}}=i(-)^{i_x} t\tau^0\\
    &u_{i,i}=\Delta \tau^1f\\
    &u_{i,i\pm2\hat{x}\pm2\hat{y}}=t^\prime \tau^3
\end{aligned}
\end{equation}
Here, we defined $\psi_i=(c_{i,\uparrow},c^\dagger_{i,\downarrow})^T$\cite{Wen2002} to write $H_{bulk}$ in a more compact form. Parameters are chosen as $t=0.5,\Delta=2,t^\prime=0.5$ to obtain a gapped $Z_2$ spin liquid. 

For the quadratic mean-field ansatz of fermionic partons, symmetries are realized projectively by a PSG\cite{Wen2002}. More precisely, each symmetry is implemented together with a gauge transformation on the partons
\begin{equation}\label{eq.3}
    \begin{aligned}
    &G_{\mathcal{T}}(x,y)=(-)^{x+y}\\
    &G_{T_x}(x,y)=(-)^y,\quad G_{T_y}(x,y)=1\\
    &G_{M_x}(x,y)=(-)^{x+y},\quad G_{M_y}(x,y)=(-)^y 
    \end{aligned}
\end{equation}
where $(x,y)\in\mathbb{Z}^2$ labels a site on the square lattice. It is easy to verify that
\begin{equation}\label{eq.1}
    M_xM_yM_x^{-1}M_y^{-1}=-1
\end{equation} 
This implies a nontrivial fractionalization class (or PSG) of the mirror symmetries $M_{x,y}$ for fermionic spinons $\epsilon$ in the $Z_2$ spin liquid, which plays an important role in the Kondo effect. 

For comparison, we can construct another model with a trivial PSG (fractionalization class) of mirror symmetries for fermionic spinons. Its Hamiltonian has the same form as Eq. \ref{Eq.bulk} and \ref{Eq.imp}, but with a different bulk mean field ansatz
\begin{equation}\label{eq.5}
\begin{aligned}
    &u_{i,i+\hat{x}}=t_1\tau^3\\
    &u_{i,i+\hat{y}}=t_2 \tau^2 +t_1\tau^3\\
    &u_{i,i}=\Delta \tau^1\\
    &u_{i,i\pm\hat{x}\pm\hat{y}}=t^\prime \tau^3
\end{aligned}    
\end{equation}
Its PSG is trivial for all space group symmetries. The following parameter set $t_1=t_2=0.5,t^\prime=0.2,\Delta=0.5$ yields a gapped $Z_2$ spin liquid. 

Next, we consider adding the impurity spin and how it is coupled to the bulk spin liquid. We make use of the $Sp(2N)$ model discussed in Ref.\cite{ran2006continuous} with $SU(2)$ gauge structure. This non-Abelian gauge structure makes it possible for the partons on the impurity site to satisfy PSG relation \ref{eq.1}. The $Sp(2N)$ impurity spin is represented by $N$ flavor spin $1/2$ fermions $f_\alpha^a$ with the following constraint
\begin{equation}\label{eq.cst}
\begin{aligned}
    &f^{a\dagger}_\alpha f^a_\alpha =N\\
    &f^a_\alpha f^a_\beta \epsilon_{\alpha,\beta} =0\\
    &f^{a\dagger}_\alpha f^{a\dagger}_\beta \epsilon_{\alpha,\beta} =0
\end{aligned}
\end{equation}
where $a=1\cdots N$ labels the flavor index and $\alpha=\uparrow,\downarrow$ labels the spin index. Repeated indices are summed over in Eq. (\ref{eq.cst}) following the Einstein convention. $\epsilon$ is antisymmetric tensor.

The coupling between the impurity and bulk $Z_2$ spin liquid is 
\begin{equation}\label{Eq.imp}
H_{imp}=\sum_{\langle j,imp \rangle} \frac{J}{N} \mathbf{S}_{j}^{a b} \cdot \mathbf{S}_{imp}^{b a}+\frac{J^\prime}{N^3} (\mathbf{S}_{j}^{a b} \cdot \mathbf{S}_{imp}^{b a})^2
\end{equation}
where $sp(2N)$ spin $\mathbf{S}$ ion the bulk is defined as\cite{ran2006continuous}
\bea
S^{ab+}=\frac12(c^{a\dagger}_\uparrow c_{\downarrow}^{b}+c^{b\dagger}_\uparrow c_{\downarrow}^{a}),~~~S^{ab,z}=\frac12(c^{a\dagger}_\uparrow c_{\uparrow}^{b}-c^{b\dagger}_\downarrow c_{\downarrow}^{a})
\eea
and the impurity spin ${\bf S}_{imp}$ is defined as 
\bea
S_{imp}^{ab+}=\frac12(f^{a\dagger}_\uparrow f_{\downarrow}^{b}+f^{b\dagger}_\uparrow f_{\downarrow}^{a}),~~~S_{imp}^{ab,z}=\frac12(f^{a\dagger}_\uparrow f_{\uparrow}^{b}-f^{b\dagger}_\downarrow f_{\downarrow}^{a})
\eea
$\langle j,imp \rangle$ denotes the 4 nearest neighbors $j$ to the impurity site, i.e. impurity is on the center of plaquette with those 4 spins on the corner, as shown in Fig. \ref{fig.model}. 

Here, we introduce the $J^\prime$ term to address a technical issue. With only the $J$ term, the mean field saddle points will be highly degenerate, which means infinitely many different mean field ansatz share the same ground state energy, an artifact of the $J$ only mean-field theory. The introduction of $J^\prime$ term reduces this infinite degeneracy to a finite fold. This technique is introduced in Ref.\cite{PhysRevB.39.11538}. Following the standard procedure, we first perform a Hubbard-Stratonovich transformation on the $J^\prime$ term and write Hamiltonian in the basis of fermions as\cite{ran2006continuous}
\begin{equation}
\begin{aligned}
H_{imp}=-\sum_{\langle j,imp \rangle} \frac{J(1+2\Phi_j J^\prime/J)}{4 N}\left(\hat{\eta}_{j,imp}^{a a \dagger} \hat{\eta}_{j,imp}^{b b}+\hat{\chi}_{j,imp}^{a a \dagger} \hat{\chi}_{j,imp}^{b b}\right) -J^\prime \Phi_j^2
\end{aligned}
\end{equation}
Where,
\begin{align}
    &\hat{\chi}_{j,imp}^{a a} = c_{j\uparrow}^{a\dagger} f_\uparrow^{a}+c_{j\downarrow}^{a\dagger} f_\downarrow^{a}\\
    &\hat{\eta}_{j,imp}^{a a} = c_{j\uparrow}^{a} f_\downarrow^{a}-c_{j\downarrow}^{a} f_\uparrow^{a}
\end{align}
Then, we do Hubbard-Stratonovich transformation to get mean field quadratic Hamitonian and ignore fluctuation of mean field, the Hamiltonian can be written as 
\begin{equation}
\begin{aligned}
H_{imp}=&\sum_{\langle j,imp \rangle} -J^\prime N \Phi_j^2 +\frac{J}{4 }( N |{\eta}_{j,imp}|^2+N|{\chi}_{j,imp}|^2 )\\
&-\frac{J}{4} \sqrt{1+2 \Phi_j J^\prime/J}(\eta_{j,imp} \hat{\eta}^{aa\dagger}_{j,imp}+\chi_{j,imp} \hat{\chi}^{aa\dagger}_{j,imp}+h.c.)
\end{aligned}
\end{equation}
Note that our convention is a bit different from Ref.\cite{PhysRevB.39.11538}. For simplicity, we define  $\tilde{J}=J \sqrt{1+2 \Phi_j J^\prime/J}$. Because of the constraint (\ref{eq.cst}) on the impurity site, we add Lagrangian multipliers $a^{\pm}=a^1\pm ia^2$ and $a_3$. Since the fluctuation can be ignored in the large $N$ limit, we have
\begin{equation}
    \begin{aligned}
    H_{imp}=
    &\sum_{\langle j,imp \rangle} -J^\prime N \Phi_j^2 +\frac{J}{4 }( N |{\eta}_{j,imp}|^2+N|{\chi}_{j,imp}|^2) \\
    &-\frac{\tilde{J}}{4} (\eta_{j,imp} (c_{j\uparrow}^{a\dagger} f_\downarrow^{a\dagger}-c_{j\downarrow}^{a\dagger} f_\uparrow^{a\dagger})
    +\chi_{j,imp}(c_{j\uparrow}^{a\dagger} f_\uparrow^{a}+c_{j\downarrow}^{a\dagger} f_\downarrow^{a})+h.c.)\\
    &+a^+ f^{a}_\uparrow f^{a}_\downarrow + a^- f^{a\dagger_\uparrow} f^{a\dagger}_\downarrow +\frac12 a^3 (f^{a\dagger}_\uparrow f^{a}_\uparrow + f^{a\dagger}_\downarrow f^{a}_\downarrow)\\
    =&\sum_{\langle j,imp \rangle}-J^\prime N \Phi_j^2+ \frac{J}{8} N \ \mathbf{tr}(u_{j,imp}^\dagger u_{j,imp})\\
    &+\frac{\tilde{J}}{4}(\psi_j^\dagger u_{j,imp} \psi_{imp}+h.c)+\psi_{imp}^{a\dagger} a_i\tau^i \psi_{imp}^a.
    \end{aligned}
\end{equation}
Here, we use the notation in Ref.\cite{Wen2002}
\begin{equation}
u_{j,imp}=\left(\begin{array}{cc}
\chi_{j,imp}^{\dagger} & \eta_{j,imp} \\
\eta_{j,imp}^{\dagger} & -\chi_{j,imp}
\end{array}\right)
\end{equation}
Combined with the bulk Hamiltonian (\ref{Eq.bulk}), we have a quadratic Hamiltonian of fermionic partons. We need to find the mean field saddle point solution for $\chi,\eta$ and $\vec a$ that extremizes the free energy. In this case, a saddle point is found where the energy is the minimum w.r.t. $\chi, \eta$ and the maximum w.r.t. $\vec a$ and $\Phi$. We obtained the saddle point solution by the gradient descent method. 

\subsection{Large-$N$ mean field solutions}

Now we discuss the results of our large $N$ mean-field calculations. The result is not sensitive to parameter $J^\prime$ and we fixed it to be $J^\prime=1.2 J$. The realistic situation of a spin-$1/2$ impurity added to a square lattice of spin-$1/2$'s is just the $N=1$ case of the $Sp(2N)$ model solved here. The main result of interest here, i.e. a symmetry protected non-Kramers doublet localized at the impurity site, is qualitatively the same for different $N\in2\mathbb{Z}+1$. Therefore we expect the same observable features of large $N$ results to also apply to the more realistic situation.

  \begin{figure}[h]
    \centering
    \includegraphics[width= 0.48\columnwidth]{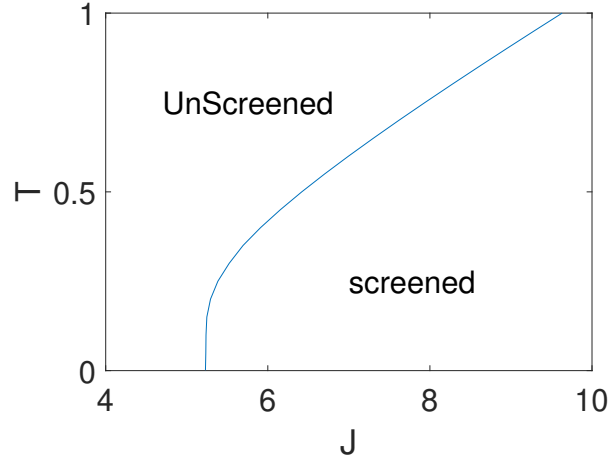}
    \caption{The phase diagram of isolated Kondo impurities in a $Z_2$ spin liquid with a nontrivial PSG. There is a Kondo temperature $T_K(J)$, denoted by the blue line, which separates the unscreened phase at high temperatures and the Kondo screening phase at low temperatues.}
    \label{fig:1}
\end{figure}

As shown in Fig. \ref{fig:1}, the $J-T$ phase diagram features two phases separated by a Kondo temperature $T_K(J)$: (i) the unscreened phase where the magnetic impurity behaves as a free magnetic moment, at a high temperature $T>T_K(J)$, and (ii) the Kondo screening phase where the magnetic impurity is screend by spinons, at a low temperature $T<T_K(J)$. Note that the Kondo screening phase only happens when the Kondo coupling $J$ is larger than a finite threshold $J_c\approx5$, as expected for Kondo effects in a gapped bulk\cite{Withoff1990,Satori1992,Saso1992,Itoh1993,Takegahara1993,Chen1998}. The criteria for the Kondo screening is the magnitude of mean-field parameters $|\chi|$ and $|\eta|$. When both parameters are zero, there is a free spin at the impurity site, pointing to the unscreened phase. When the mean-field parameters become nonzero, the impurity becomes a part of the bulk spin liquid, with a gapped spectrum for the full Hamiltonian. This corresponds to the Kondo screening phase. 

\begin{figure}[h]
    \centering
    \includegraphics[width= 0.48\columnwidth]{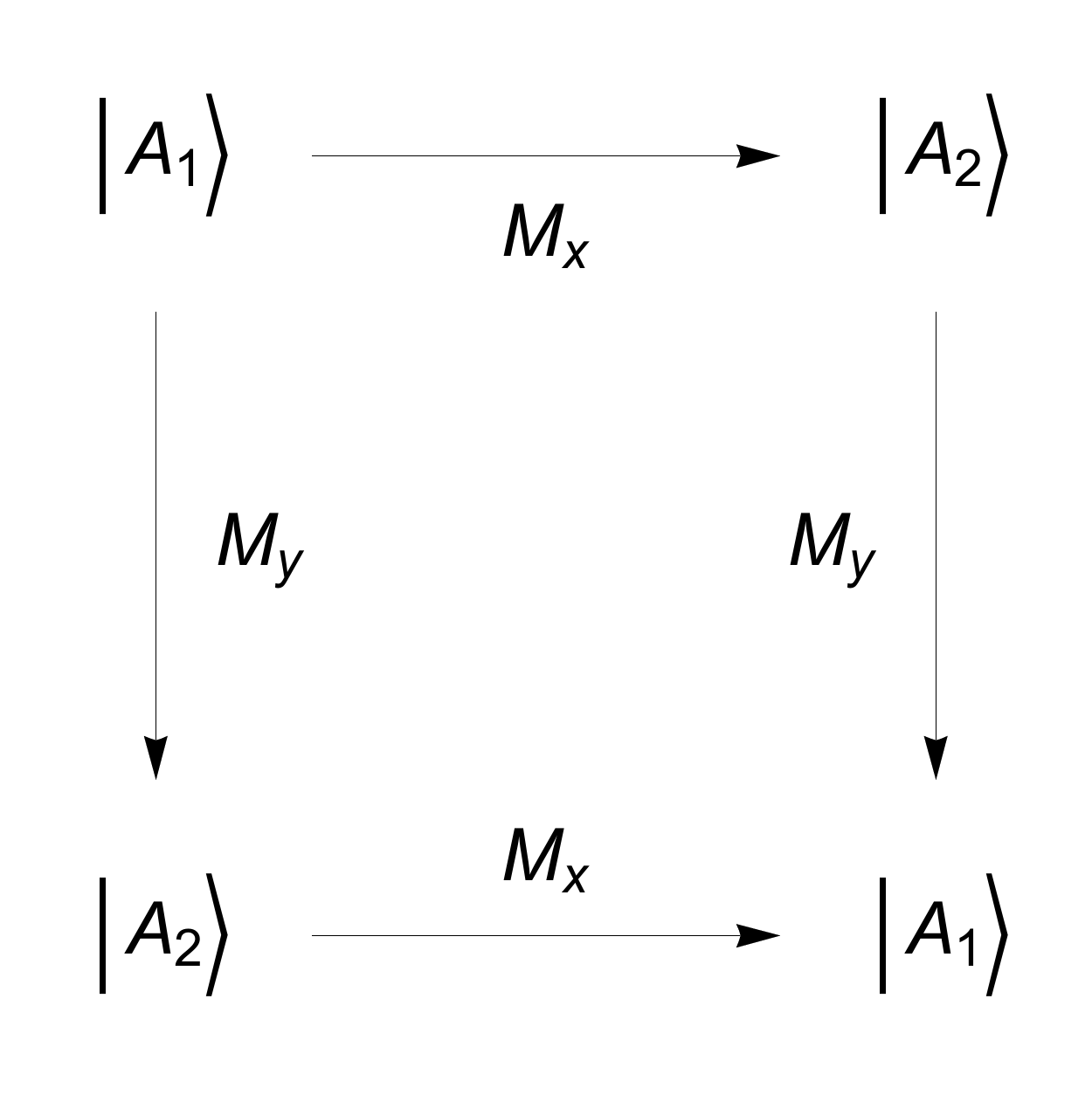}
    \caption{Illustration of symmetry acting on ground states. $|A_1\rangle,|A_2\rangle$ mean two states corresponding to two mean-field ansatz.}
    \label{fig:8}
\end{figure}

In the Kondo screening phase, we obtain two saddle point solutions that minimize the free energy:
\begin{equation}
\begin{aligned}
    &u_{(1,1),imp}=u_1 i\tau^0\\
    &u_{(2,1),imp}=\frac{u_1}{\sqrt{2}} i\tau^0-\frac{u_1}{\sqrt{2}} \tau^1\\
    &u_{(1,2),imp}=-\frac{u_1}{\sqrt{2}} i\tau^0-\frac{u_1}{\sqrt{2}} \tau^1\\
    &u_{(2,2),imp}=u_1 \tau^1\\
    &a= a_1 \tau^1
\end{aligned}
\end{equation}
and
\begin{equation}
\begin{aligned}
    &u_{(1,1),imp}=-\frac{u_1}{\sqrt{2}}\tau^2+\frac{u_1}{\sqrt{2}} \tau^3\\
    &u_{(2,1),imp}=u_1 \tau^2\\
    &u_{(1,2),imp}=u_1 \tau^3\\
    &u_{(2,2),imp}=\frac{u_1}{\sqrt{2}}\tau^2+\frac{u_1}{\sqrt{2}} \tau^3\\
    &a= -a_1 \tau^1
\end{aligned}
\end{equation}
where $(x,y)$ with $x,y=1,2$ labels the four neighboring spins closest to the impurity. $u_1$ and $a_1$ are two positive numbers that are functions of $J$ and $T$. We label there two mean-field ansatz $A_1$ and $A_2$. They are related by mirror symmetries $M_x$ and $M_y$, as illustrated in Fig. \ref{fig:8}. The gauge transformation associated with $M_x$ and $M_y$ in the bulk are shown in Eq. (\ref{eq.3}). On the impurity site, the gauge transformations associated with the two mirrors are $G_{M_x}= i\tau^2$ and $G_{M_y}=-i\tau^3$.

On a square lattice of $N^2$ sites ($N$= even) and 1 impurity spin, with an odd number of fermions due to on-site constraint (\ref{eq.cst}), due to the nontrivial PSG in (\ref{eq.2}), $M_x$ and $M_y$ operators anticommute with each other in the two mean-field ground states $|A_1\rangle$ and $|A_2\rangle$\cite{Lu_2018}. Since the two mean-field states are related by mirror symmetry $M_x$ (or $M_y$), as shown in Fig. \ref{fig:8}, the representation of the two mirror symmetries in this two-dimensional space must have the following form in the basis of $(|A_1\rangle,|A_2\rangle)^T$:
\bea
R(M_x)=\sigma_x,~~~R(M_y)=e^{i\phi}\sigma_y
\eea
by a proper gauge choice, where $\sigma_{x,y,z}$ are the Pauli matrices. As a result, any local perturbation $\hat H$ preserving $M_{x,y}$ symmetries must have the following form:
\bea
\langle A_i|\hat H|A_j\rangle=h\delta_{i,j}
\eea
In other words, the two mean-field states form a projective representation of the impurity site symmetry group $G_s=Z_2^{M_x}\times Z_2^{M_y}$, analogous to the exactly solvable model discussed previously in section \ref{section.1}. This means the 2-fold degeneracy between saddle point solutions $|A_1\rangle$ and $|A_2\rangle$ cannot be split by any symmetry-preserving local perturbations. For example, we have explicitly checked that in our case 
\begin{equation}\label{eq.6}
    \langle A_1 | \sum_{\langle j,imp \rangle} \vec{S}_j \cdot \vec{S}_{imp} | A_2 \rangle\equiv0
\end{equation}
As a result, for a $Z_2$ spin liquid with a nontrivial symmetry fractionalization class, the Kondo screening phase is characterized by a symmetry protected non-Kramers doublet at the impurity site.

To compare, we also consider the case of a bulk $Z_2$ spin liquid with a trivial PSG, shown in Eq. (\ref{eq.5}), where $M_x$ and $M_y$ commute with each other when acting on a single parton operator. The unscreened phase is still featured by $\chi=\eta=0$, leaving a free spin at the impurity site. Meanwhile, in the Kondo screening phase, there is a unique mean-field ansatz as a gapped symmetric ground state. The associated mean-field ansatz is 
\begin{equation}
\begin{aligned}
    &u_{(1,1),imp}=u_1 i\tau^0\\
    &u_{(2,1),imp}=u_1 i\tau^0\\
    &u_{(1,2),imp}=-u_1 i\tau^0\\
    &u_{(2,2),imp}=-u_1 i\tau^0\\
    &a= a_1 \tau^1 + a_2\tau^2 +a_3 \tau^3
\end{aligned}
\end{equation}
where parameters $u_1,a_i,i=1\cdots 3$ are functions of $T,J$. This mean-field ansatz is invariant under both mirror symmetries as we choose gauge transformations $G_{M_x}=1,G_{M_y}=-1$ on the impurity site. This unique paramagnetic ground state distinguishes itself from the 2-fold degeneracy in the case of a nontrivial PSG, because here there is no local degeneracy protected by mirror symmetries.

\subsection{Thermodynamic quantities}

We perform a self-consistent mean-field calculation in the large $N$ limit, on a square lattice of $20 \times 20$ sites. To avoid an extra spinon lurking in the bulk, we consider two magnetic impurities separated by a largest distance ($d=10\sqrt2$) between each other.

\begin{figure}
    \centering
    \includegraphics[width= \columnwidth]{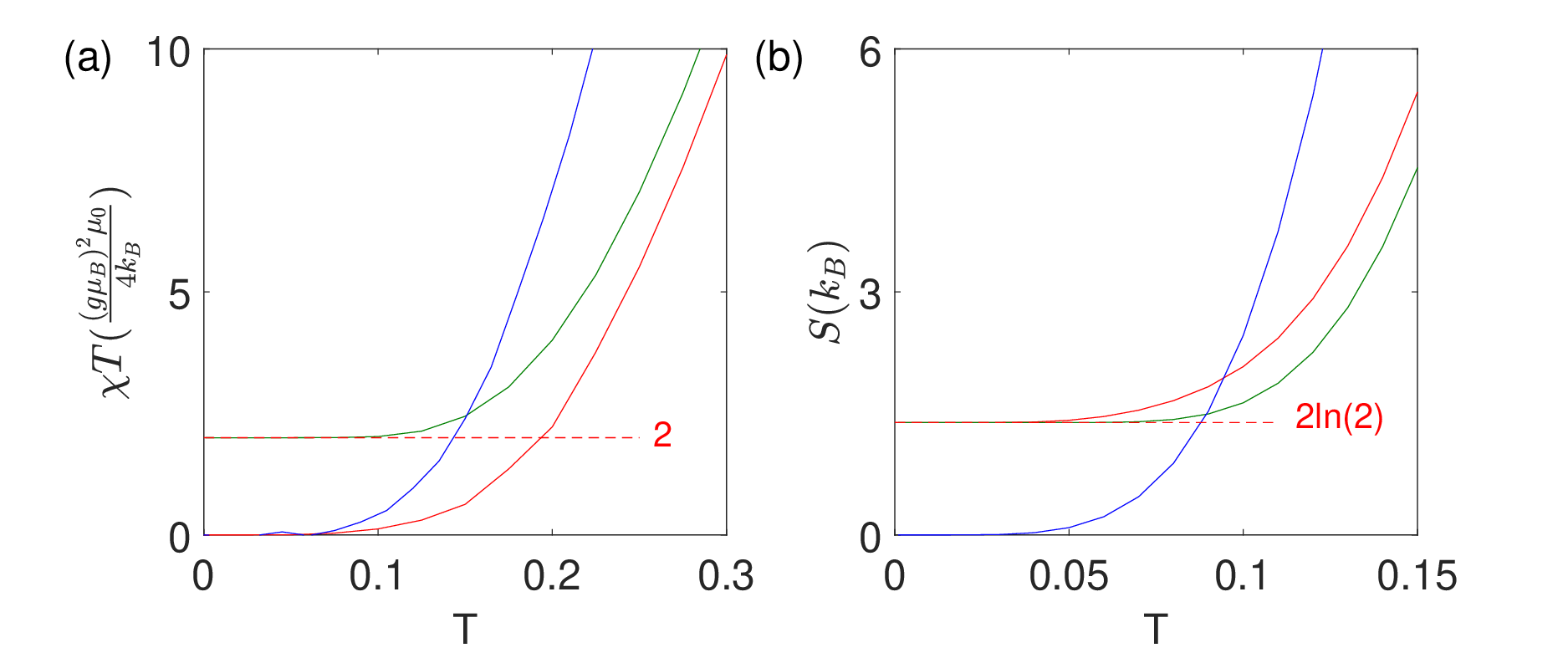}
    \caption{ The temperature dependence of (a) uniform magnetic susceptibility $\chi(T)$ and (b) entropy $S(T)$ contributions from Kondo impurities in different regimes: the unscreened regime of free moments at the impurity sites (green), the Kondo screening regime in $Z_2$ spin liquids with a trivial (blue) vs. nontrivial (red) spinon fractionalization class. The calculations are preformed on a $20\times20$ square lattice with 2 spatially separated magnetic impurities.}
    \label{fig:thermo}
\end{figure}

In Fig. \ref{fig:thermo}, we show the thermodynamic quantities from the mean-field calculation to illustrate the experimental implications of the anomalous Kondo effect discussed above. 

The uniform magnetic susceptibility $\chi(T)$ behaves as $1/T$ for free spins, and therefore $\chi\cdot T$ approaches a constant at low temperatures in the unscreened phase. In the Kondo screening regime, however, $\chi(T)$ decays exponentially as $\sim e^{-\Delta/k_BT}$ below the Kondo temperature $T_K(J)$, where $\Delta$ is the spin gap of the system. Since the ground states always feature a finite spin gap for both the case of a trivial and that of a nontrivial PSG, the susceptibility can only differentiate unscreend phase from the Kondo screening phase, but cannot differentiate a trivial PSG from a nontrivial one. 

On the other hand, the specific heat $C_v(T)$, or the entropy
\bea
S(T)=\int_0^{T}\frac{C_v(t)}{t}\text{d}t
\eea
at low temperatures can measure the density of states at a low energy, and can be used to differentiate a trivial PSG from a nontrivial one. To be precise, in the unscreened phase, each impurity site with a free moment contribute a $k_B\ln2$ at a low temperature. Meanwhile, in the Kondo screening regime, for a trivial PSG, the system is gapped everywhere, leading to an exponentially vanishing specific heat and entropy. This is in sharp contrast to a nontrivial impurity, where each impurity site features a symmetry-protected 2-fold non-Kramers doublet, and contributes a $k_B\ln2$ entropy at low temperatures.

In reality, in a quasi-2d material with a dilute concentration of impurities satisfying $\rho\xi^2\ll1$, where $\rho$ is the impurity density and $\xi$ is the correlation length, we expect the observable signatures discussed above to remain valid. In particular, there will be a low temperature peak in the specific heat at $T\sim Je^{-C_0/\xi\sqrt{\rho}}$, where $C_0$ is a constant of order one. The entropy plateau will persist between this low temperature scale and the Kondo temperature $T_K(J)$.

\end{document}